\documentclass[preprint,11pt]{elsarticle}

\usepackage{amsmath,amssymb,amsfonts}
\usepackage{url}
\usepackage{graphicx}
\usepackage{textcomp}
\usepackage{hyperref}
\usepackage{multicol}
\usepackage{multirow}
\usepackage[table,xcdraw]{xcolor}
\usepackage{xcolor}
\usepackage{adjustbox}
\usepackage{ragged2e}
\usepackage{array}
\usepackage{mathtools}
\usepackage{rotating}
\usepackage{lipsum}
\usepackage{subfigure}
\usepackage[utf8]{inputenc}
\usepackage{booktabs}
\usepackage{adjustbox}
\usepackage{tabularx}
\usepackage[many]{tcolorbox}

\usepackage{amsthm}%
\usepackage{mathrsfs}%
\usepackage[title]{appendix}%
\usepackage{xcolor}%
\usepackage{textcomp}%
\usepackage{manyfoot}%
\usepackage{booktabs}%
\usepackage{url}
\usepackage{algpseudocode}%
\usepackage{listings}%
\usepackage{tabularx}
\usepackage{caption}
\usepackage[many]{tcolorbox}
\usepackage{booktabs}
\usepackage{pifont}

\usepackage{booktabs}
\usepackage{fontawesome5} 
\usepackage{xcolor}  
\newtcolorbox{boxK}{
    sharpish corners, 
    boxrule = 0pt,
    toprule = 0.5pt, 
    enhanced,
    fuzzy shadow = {0pt}{-2pt}{-0.5pt}{0.5pt}{black!35} 
}

\usepackage{geometry}
\journal{Journal of Systems and Software}

\begin{document}

\begin{frontmatter}

\title{From Prompt to Pipeline: Large Language Models for Scientific Workflow Development in Bioinformatics}

\author{Khairul Alam}
\author{Banani Roy}          
\affiliation{organization={Department of Computer Science, University of Saskatchewan},
            addressline={110 Science Pl}, 
            city={Saskatoon},
            postcode={S7N 5A2}, 
            state={SK},
            country={Canada}}
\begin{abstract}
Scientific Workflow Systems (SWSs) such as Galaxy and Nextflow are essential for scalable, reproducible, and automated bioinformatics analyses. However, developing and understanding scientific workflows remains challenging for many domain scientists due to the complexity of tool/module selection, infrastructure requirements, and limited programming expertise. This study explores whether state-of-the-art Large Language Models (LLMs) such as GPT-4o, Gemini 2.5 Flash, and DeepSeek-V3 can assist in generating accurate, complete, and usable bioinformatics workflows. We evaluate a set of representative workflows covering tasks such as RNA-seq, SNP analysis, and DNA methylation across both Galaxy (graphical) and Nextflow (script-based) platforms. To simulate realistic usage, we adopt a tiered prompting strategy: each workflow is first generated using an instruction-only prompt; if the output is incomplete or incorrect, we escalate to a role-based prompt, and finally to chain-of-thought prompting if needed. The generated workflows are evaluated against community-curated baselines from the Galaxy Training Network (GTN) and nf-core, using criteria including correctness, completeness, tool appropriateness, and executability. Results show that LLMs exhibit strong potential in workflow development. Gemini 2.5 Flash produced the most accurate and user-friendly workflows in Galaxy, while DeepSeek-V3 excelled in Nextflow pipeline generation. GPT-4o performed nicely with structured prompts. Prompting strategy significantly influenced output quality, with role-based and chain-of-thought prompts enhancing correctness and completeness. Overall, LLMs can reduce the cognitive and technical barriers to workflow development, making SWSs more accessible to novice and expert users. This work highlights the practical utility of LLMs and provides actionable insights for integrating them into real-world bioinformatics workflow design.
\end{abstract}



\begin{keyword}
Scientific Workflows \sep Scientific Workflow Systems \sep Large Language Models \sep Bioinformatics \sep Prompting Techniques \sep Natural Language to Workflow
\end{keyword}

\end{frontmatter}

\section{Introduction}\label{introduction}
The exponential growth of biological data and the increasing complexity of computational analyses have made scientific workflows \cite{liu2015survey} indispensable in modern bioinformatics \cite{cohen2017scientific}. These workflows (workflow means scientific workflow for the rest of the paper) are essential for enabling scalable, reproducible, and dependable large-scale data analysis, particularly when deployed on distributed and cloud infrastructures \cite{sanger2024qualitative, wratten2021reproducible}. In bioinformatics, workflows routinely process terabyte-scale datasets generated by advanced DNA, RNA, NGS sequencing, and other technologies \cite{wratten2021reproducible} across diverse experimental contexts \cite{goodwin2016coming}. Typically, these workflows comprise multiple computational steps, including data preprocessing, quality control, sequence aggregation, machine learning for classification or clustering, statistical analysis, and result visualization. Each step is carried out by specific tools/modules, often developed by third parties, facilitated by scientific workflow systems (SWSs)' no/low-code environments \cite{xu2024llm4workflow}. SWSs have become essential platforms for constructing, executing, and managing bioinformatics workflows by integrating diverse analytical and data processing tools/modules into reproducible pipelines (workflows) \cite{langer2024empowering}. SWSs such as Galaxy \cite{goecks2010galaxy}, Nextflow \cite{di2017nextflow}, Snakemake \cite{koster2012snakemake}, Airflow \cite{harenslak2021data}, and others \cite{cohen2017scientific, wratten2021reproducible, deelman2015pegasus, goodstadt2010ruffus} provide structured environments that support the automation, customization, and reuse of complex analyses. In response to the growing demand for scalable and transparent data processing, hundreds of SWSs \cite{sws-list} have emerged, offering access to thousands of bioinformatics tools/modules. For instance, Galaxy’s ToolShed \footnote{https://toolshed.g2.bx.psu.edu/} alone hosts over 10,600 tools,  Nextflow has over 1550 modules \footnote{https://nf-co.re/modules/}, enabling researchers to assemble rich and versatile workflows tailored to their specific needs. These platforms have significantly advanced reproducibility and efficiency in computational biology by facilitating workflow sharing, versioning, and collaboration \cite{kumar2021tool}.

Despite the widespread adoption of SWSs, workflow development remains a significant challenge for both novice and expert users \cite{alam2023reusability, ferreira2024workflows, menaka2022workflow, deelman2018future}. Practitioners, particularly those without programming backgrounds, face steep learning curves due to the need for scripting, tool configuration, dependency management, and infrastructure provisioning \cite{alam2025empirical}. Workflow developers are usually domain scientists with deep subject-matter expertise but limited experience in software engineering or distributed/cloud computing. While SWSs offer a broad array of services, such as data cleaning, statistical analysis, modeling, and high-performance computing, these capabilities are typically leveraged effectively only by a minority of users who are proficient in workflow composition, programming, and infrastructure management \cite{de2007myexperiment}. Consequently, many researchers struggle to design workflows to manage complex data and operate on these SWSs. These difficulties stem from the diversity of tools/modules, scripts, and data formats essential to ensure data integrity, reproducibility, and scalability. The resulting complexity hampers workflow development, adaptation, and interpretation, ultimately impeding data exploration and scientific discovery \cite{deelman2018future, cohen2011search}.

Recent advancements in LLMs such as GPT-4o \cite{hurst2024gpt}, Gemini 2.5 Flash \cite{comanici2025gemini}, and DeepSeek-v3 \cite{liu2024deepseek} have shown strong capabilities in understanding complex instructions and generating accurate, context-aware code, making them highly relevant for code generation and software development tasks \cite{ouyang2023llm, hou2023large}. However, their application in real-world bioinformatics workflow development remains underexplored. This gap underscores the need to assess how LLMs can support practitioners in developing scientific workflows. Given their code generation abilities, we posit that LLMs can produce accurate workflows with appropriate prompting. We investigate this potential using three state-of-the-art LLMs. GPT-4o offers broad tool knowledge and excels in general-purpose coding tasks; Gemini 2.5 Flash demonstrates strengths in structured reasoning and system integration; and DeepSeek-v3, a code-focused open-source model, provides a valuable perspective beyond proprietary alternatives.

To evaluate the capabilities of LLMs in assisting workflow development, we focus on two widely adopted and most popular SWSs for the bioinformatics domain: \emph{Galaxy} \cite{goecks2010galaxy} and \emph{Nextflow} \cite{di2017nextflow}. These platforms represent complementary paradigms; Galaxy offers a user-friendly, graphical interface tailored for domain scientists with limited programming experience, while Nextflow provides a script-based environment optimized for developers seeking scalability and reproducibility. Galaxy is known for its accessibility, extensive tool repository via ToolShed, and emphasis on transparency and reproducibility through automatic logging of analysis steps. It supports deployment across diverse infrastructures, including local servers, institutional clusters, and cloud platforms. In contrast, Nextflow enables the composition of complex, modular pipelines inspired by Unix principles, with built-in support for containerization (Docker, Singularity) and distributed and cloud computing environments (e.g., Kubernetes, AWS Batch, Slurm). Its portability and integration with version control and software packaging tools ensure the principle of \emph{write once, run anywhere}. Together, Galaxy and Nextflow provide a robust foundation for assessing the generalizability, usability, and automation potential of LLMs in workflow development across both GUI-driven and script-driven environments.

Integrating LLMs into bioinformatics workflow development presents a compelling opportunity to reduce the cognitive and technical burden on workflow developers. With appropriate prompting, LLMs could potentially help users generate workflows, configure parameters, select tools/modules, and even interpret error messages. However, concerns persist regarding the reliability of generated outputs, especially in scientific contexts where precision and reproducibility are critical. To ensure the effectiveness of our approach, we have meticulously applied state-of-the-art prompt engineering strategies, drawing on insights from prominent studies in this field \cite{ekin2023prompt, arora2022ask, reynolds2021prompt, wei2022chain, zhou2022large, marvin2023prompt, lin2024write}.

In this study, we conduct a comprehensive evaluation of state-of-the-art LLMs (\emph{GPT-4o, Gemini 2.5 Flash, and DeepSeek-V3}) in their ability to support workflow development in bioinformatics. We begin by assessing the capacity of these models to explain foundational concepts related to scientific workflows and SWSs, followed by an in-depth evaluation of their ability to develop workflows across two widely used platforms: \emph{Galaxy} and \emph{Nextflow}. We then analyze the generated workflows for correctness, completeness, and usability, comparing them against community-curated baselines (Galaxy Training Network \cite{hiltemann2023galaxy} and nf-core \cite{ewels2020nf}). Furthermore, we investigate how different prompting strategies (e.g., instruction-only, role-based, and chain-of-thought) affect output quality. By identifying recurring challenges and model-specific limitations, we propose directions for improving LLM-supported workflow development, including integrating domain-specific knowledge and refining prompt engineering techniques. This work contributes to the emerging intersection of natural language processing and computational bioinformatics by offering practical insights into how LLMs can bridge the gap between domain scientists and automated workflow generation. Specifically, we aim to answer the following three research questions (RQs):

\begin{itemize}
    \item \textbf{RQ1: }To what extent can LLMs (e.g., GPT-4o, Gemini 2.5 Flash, and DeepSeek-V3) assist in developing scientific workflows?
    \item \textbf{RQ2: }How do workflows generated by different LLMs compare regarding completeness, correctness, and usability?
    \item \textbf{RQ3: }What prompting strategies should a workflow developer follow? 
\end{itemize}

\textbf{Paper Organization: }The remainder of this paper is outlined as follows: Section \ref{background-of-prompts-to-pipeline} provides background on Scientific Workflows, Scientific Workflow Systems (SWSs), Large Language Models (LLMs), and prompting techniques, establishing the foundation for our investigation. Section \ref{study-design-of-prompts-to-pipeline} outlines our study design, including the selection of workflow platforms, LLMs, and evaluation criteria. In Section \ref{results-of-prompts-to-pipeline}, we present the findings of our analysis, followed by an in-depth discussion and interpretation in Section \ref{prompt-to-pipeline-discussion}. Section \ref{threats-of-prompts-to-pipeline} examines potential threats to the validity of our results, considering both methodological and contextual limitations. Section \ref{related-work-of-prompts-to-pipeline} situates our research within the broader landscape of related efforts. Finally, Section \ref{conclusion-of-prompts-to-pipeline} concludes the paper, highlighting directions for future research.

\section{Background} \label{background-of-prompts-to-pipeline}
Scientific workflows combine multiple software artifacts on distributed stacks for advanced data analysis \cite{sanger2024qualitative}. We ground the reader by providing a brief overview of scientific workflows and workflow systems and introducing LLMs, including their utility to facilitate the creation of workflows.

\subsection{Scientific Workflows}
\begin{figure*}[htbp]
    \vspace{-0.5em}
    \centering
    \includegraphics[width=\textwidth]{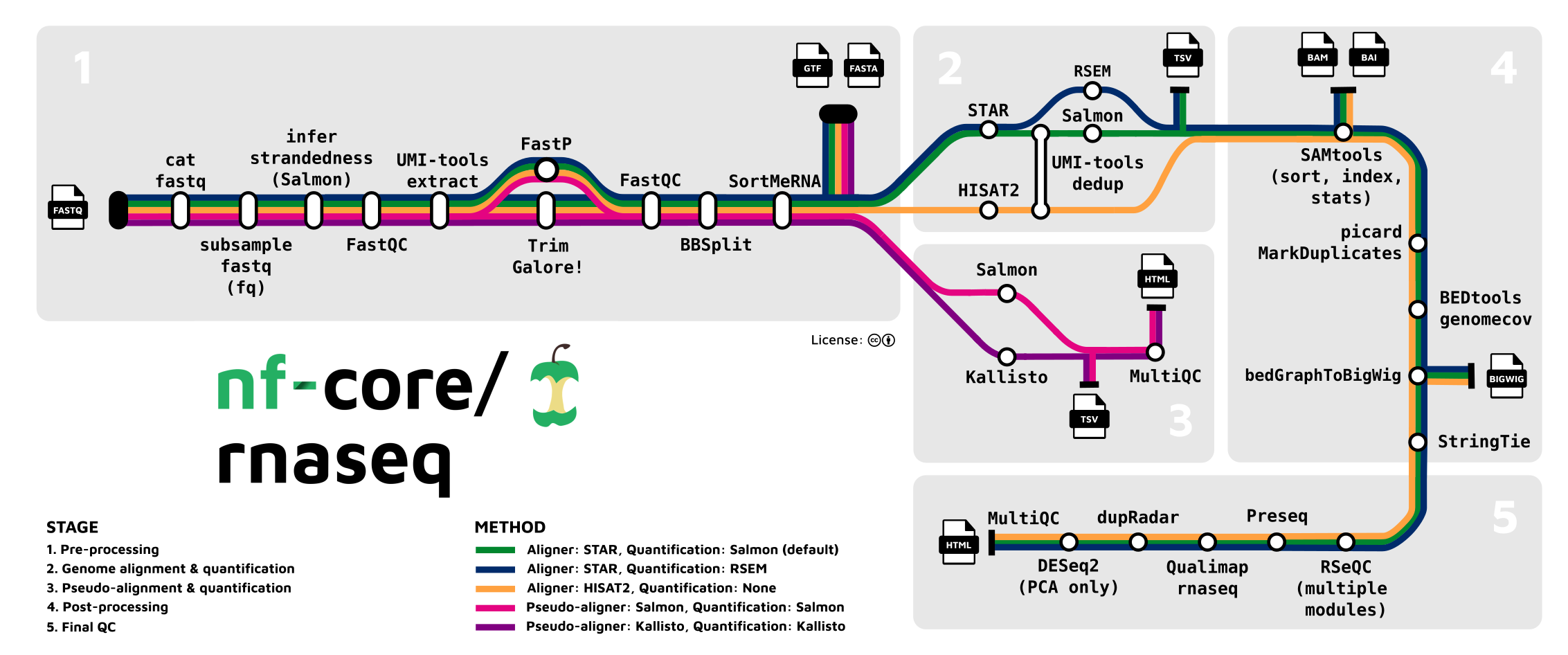}
    \caption{RNA sequencing analysis pipeline (workflow)}
    \label{fig:rnasequencing}
    \vspace{-1em}
\end{figure*}

A scientific workflow is a structured and reproducible sequence of data processing tasks, typically modeled as a directed acyclic graph (DAG), where nodes represent computational operations and edges indicate data dependencies \cite{liu2015survey, barker2007scientific}. It automates complex scientific procedures, such as data acquisition, transformation, and analysis, thereby accelerating discovery through efficient and reliable execution \cite{ludascher2009scientific}. Analogous to a cooking recipe, where ingredients, preparation steps, and instructions combine to produce a consistent outcome, scientific workflows ensure reproducibility and scalability in computational experiments. In this paper, we use the terms \emph{workflow, scientific workflow, and pipeline} interchangeably to refer to scientific workflow. A workflow can be graphic-based and script-based. Figure~\ref{fig:rnasequencing} illustrates an example of an \emph{RNA sequencing} workflow steps developed using \emph{Nextflow} SWS, obtained from \cite{ewels2020nf}. This pipeline takes FASTQ files as input, performs quality control and trimming, aligns or pseudo-aligns reads to the reference genome, quantifies gene and transcript expression, and generates a comprehensive expression matrix alongside detailed QC reports.

\subsection{Scientific Workflow Systems}
SWSs are specialized software platforms designed to automate, manage, and execute complex sequences of computational tasks, where the execution order is controlled by a formal representation of workflow logic \cite{lin2009reference}. In bioinformatics, SWSs play a critical role in handling large-scale, heterogeneous data by enabling reproducible, scalable, and modular analysis pipelines. Prominent systems such as Galaxy \cite{goecks2010galaxy}, Nextflow \cite{di2017nextflow}, Taverna \cite{oinn2004taverna}, Snakemake \cite{koster2012snakemake}, Apache Airflow \cite{harenslak2021data}, and Pegasus \cite{deelman2015pegasus} have been developed to support domain scientists in composing and executing workflows for a wide range of bioinformatics tasks, including genome assembly, variant calling, RNA-seq analysis, and metagenomics. The increasing reliance on workflow-based analysis in computational biology has led to the emergence of hundreds of SWSs \cite{sws-list}, each offering varying degrees of abstraction and control. Core features commonly supported by these platforms include graphical or script-based workflow design, task scheduling and execution, real-time monitoring, provenance tracking for reproducibility, error handling, and integration with high-performance or cloud computing infrastructures \cite{liew2016scientific}. SWSs thus provide an essential foundation for managing the complexity and scale of modern bioinformatics research.

\subsection{Large Language Models}
LLMs are advanced AI systems capable of processing and generating human-like text using vast training data. Their development has been revolutionized by the transformer architecture introduced by Vaswani et al. in \emph{Attention Is All You Need} \cite{vaswani2017attention}. Leveraging neural networks and transformer variants, LLMs like GPT-3, GPT-4, Codex (OpenAI), DeepSeek-V3 (DeepSeek), Gemini, PaLM (Google), LLaMA (Meta), and Claude (Anthropic) and others \cite{zhao2023survey} have become cornerstones of natural language processing. Tools like Codex and ChatGPT assist developers by suggesting code snippets, completing functions, and enhancing productivity \cite{liu2024your}. They aid in code review, bug detection, and quality assurance, promoting reliable software \cite{yu2024security}. Additionally, they generate and update documentation, user guides, and API references, simplifying software maintenance \cite{nam2024using, su2023hotgpt}. LLMs even support cross-platform development by translating code between languages, improving collaboration among diverse teams \cite{pan2023understanding, yang2024exploring}. LLMs are widely acknowledged for their ability to generate commonsense knowledge, leveraging the vast training data \cite{sanger2024qualitative}. This capability makes them practical and accessible knowledge repositories that can be queried through natural language prompts. LLMs are also increasingly used in bioinformatics for tasks like protein structure prediction, genomic sequence analysis, drug discovery, gene expression profiling, and pathway analysis, leveraging their ability to model complex biological data with transformer-based architectures \cite{sarumi2024large}. A key limitation of LLMs is their tendency to produce \emph{hallucinations} \cite{manakul2023selfcheckgpt, peng2023check}.

\subsection{Usages of LLMs}
Recent advances in LLMs have driven interest in automating software development tasks such as code generation, refactoring, and debugging. Studies show that models like Codex, ChatGPT, and GitHub Copilot can generate syntactically correct and semantically meaningful code from natural language prompts \cite{ugare2024improving, gu2023llm, le2022coderl}, and also assist with interactive tasks like modifying code based on user intent or incomplete specifications \cite{jiang2022discovering}. They help reduce reliance on online searches and offer scaffolds for common tasks, though challenges persist in debugging and verifying complex outputs \cite{vaithilingam2022expectation}. Despite these issues, LLMs have proven effective in many software development contexts \cite{yeticstiren2023evaluating, kazemitabaar2023studying}.

While LLMs have been widely studied in the context of general software engineering, their application in scientific data analysis and workflow automation, particularly in domains like bioinformatics, remains relatively nascent. Initial efforts by \cite{liu2023wants} explored how code generators like Codex could assist non-programmers in performing data analysis tasks. Similarly, \cite{sanger2024qualitative} conducted a qualitative assessment of ChatGPT's effectiveness in supporting scientific workflow design. More recently, \cite{xu2024llm4workflow} introduced LLM4Workflow, a framework that integrates LLMs into a workflow model. LLMs like BERT and GPT-series have been increasingly adapted for bioinformatics, leading to the development of specialized models such as DNABERT \cite{ji2021dnabert}, ProteinBERT \cite{brandes2022proteinbert}, DNAGPT \cite{zhang2023dnagpt}, and ProtGPT2 \cite{ferruz2022protgpt2}. These domain-specific variants are designed to process and interpret the unique linguistic structures of biological sequences, such as DNA, RNA, and proteins by conceptualizing them as natural language. Leveraging transformer-based architectures, these models demonstrate strong performance in a variety of tasks, including gene annotation, protein function prediction, variant calling, and motif discovery. Their ability to handle large-scale, complex, and unstructured biological data makes them powerful tools for uncovering meaningful biological patterns, enhancing data-driven decision-making, and accelerating discoveries in genomics, and proteomics. Building on these foundational studies, our research investigates the underexplored potential of LLMs in the domain of bioinformatics workflow development. Specifically, we assess how state-of-the-art LLMs can generate, explain, and refine workflows for scientific tasks. By systematically analyzing the influence of different prompting strategies, such as \emph{instruction-only, role-based, and chain-of-thought}, we aim to unlock more effective and reliable uses of LLMs in enabling non-programmers and domain scientists to construct robust, reproducible, and scalable workflows. This work contributes to bridging the gap between advanced language models and practical scientific workflow development.

\subsection{Prompt Engineering}
Prompt engineering refers to the process of crafting input instructions that effectively guide an LLM to generate desired outputs. A prompt typically consists of natural language text that conveys a task, includes relevant context, and may specify output format or style \cite{marvin2023prompt}. Since LLMs are sensitive to how information is framed, the structure, clarity, and specificity of a prompt significantly influence the relevance, accuracy, and usefulness of the model's response \cite{oppenlaender2023taxonomy}. Prompt engineering has therefore emerged as a critical practice for aligning LLM behavior with user goals, especially in complex domains such as scientific workflow development. In the context of this study, prompt engineering plays a pivotal role in enabling LLMs to understand domain-specific requirements in bioinformatics and translate them into structured workflows. Drawing from recent research that emphasizes prompt design as a programming paradigm in itself \cite{white2023prompt}, we explore strategies \cite{lin2024write, liu2023jailbreaking, marvin2023prompt, arora2022ask, ekin2023prompt, reynolds2021prompt, wei2022chain, zhou2022large, wang2024prompt, sahoo2024systematic, chen2023unleashing, parameswaran2023revisiting, vatsal2024survey, shah2025prompt, smith2024language, cohn2023towards, razdaibiedina2023progressive,knoth2024ai, lo2023art} for adapting prompts to scientific tasks, such as generating analysis pipelines or interpreting bioinformatics procedures. These strategies include varying prompt templates, combining task descriptions with expected outcomes, and incorporating domain-relevant terminology to improve performance across different use cases \cite{lo2023clear}.

To systematically investigate prompt effectiveness, we evaluate three widely used prompting techniques: (1) \emph{instruction-only} prompting, where the task is explicitly described in a natural language instruction; (2) \emph{role-based} prompting, where the model is asked to assume the perspective of a domain expert; and (3) chain-of-thought prompting, which encourages step-by-step reasoning. These three techniques represent diverse and widely adopted paradigms in prompt engineering, each targeting a different axis of LLM interaction: clarity of task (instruction-only), context emulation (role-based), and reasoning process (chain-of-thought). Instruction-only prompts are the most accessible and commonly used format in practical applications, offering a strong baseline for comparison. Role-based prompts introduce a layer of contextual alignment by asking the model to simulate expert knowledge, which is particularly relevant in specialized domains like bioinformatics. Meanwhile, chain-of-thought prompting has been shown to enhance performance on tasks that require multi-step reasoning and logical coherence, traits that mirror the sequential and conditional structure of scientific workflows. Together, these three approaches span a broad spectrum of prompting complexity and cognitive alignment, allowing us to meaningfully assess the impact of prompt design on workflow generation quality. By focusing on these strategies, we aim to provide actionable insights that are both theoretically grounded and practically relevant for researchers and developers using LLMs in bioinformatics and beyond. As scientific workflows often involve multi-step reasoning, domain-specific knowledge, and structured output, prompt engineering becomes essential in bridging the gap between high-level user intent and executable computational pipelines. By comparing prompt outcomes across different LLM platforms, our study contributes empirical insights into the design of effective prompts for scientific workflow generation. This reinforces the notion that prompt engineering is not merely a preliminary step but a central component in harnessing LLMs for domain-specific scientific computing.

\section{Study Design} \label{study-design-of-prompts-to-pipeline}
\begin{figure*}[htbp]
    \vspace{-1.5em}
    \centering
    \includegraphics[width=\textwidth]{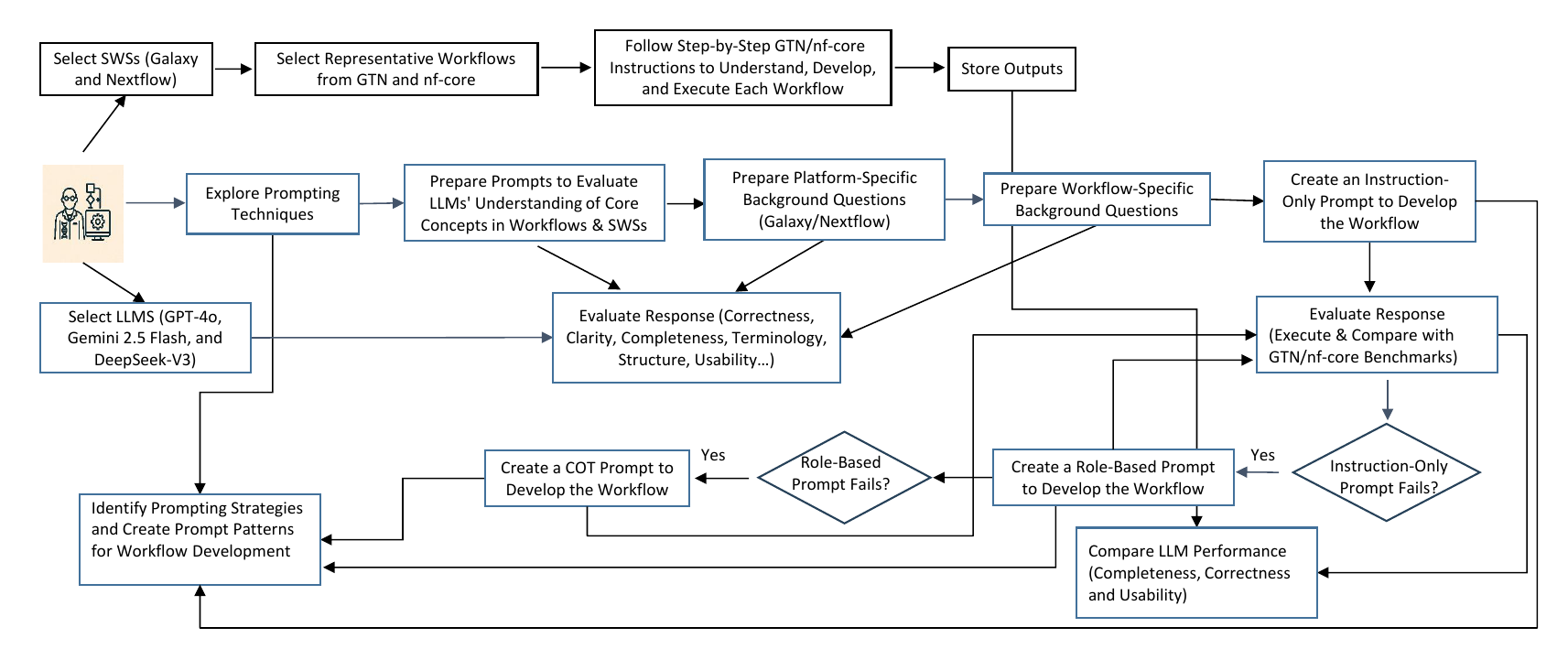}
    \caption{Overview of our study methodology}
    \label{fig:studymethodology}
        \vspace{-1.5em}
\end{figure*}

This study investigates how state-of-the-art LLMs, \emph{GPT-4o, Gemini 2.5 Flash, and DeepSeek-V3}, can generate bioinformatics workflows from natural language prompts using two widely adopted Scientific Workflow Systems: Galaxy and Nextflow. Our experimental design focuses on three main research questions: (\textbf{RQ1}) \emph{How effectively can LLMs support workflow development?}, (\textbf{RQ2}) \emph{How do their outputs compare in terms of quality?} (\textbf{RQ3}) \emph{Which prompting strategies yield the best results?}. Through this investigation, we aim to assess the technical capabilities of LLMs and identify best practices in prompt formulation that can support novice and expert users in developing robust, domain-relevant bioinformatics workflows. Figure \ref{fig:studymethodology} shows the overall approaches of our study methodology.

We select \emph{Galaxy} and \emph{Nextflow} as representative SWSs due to their widespread adoption within the bioinformatics community \cite{wratten2021reproducible, langer2024empowering} and complementary design paradigms. For model selection, we consider three advanced LLMs with diverse strengths. \emph{GPT-4o} is chosen for its demonstrated superiority in code generation and reasoning tasks across multiple domains. \emph{Gemini 2.5 Flash} is included for its efficiency in producing structured outputs and handling multi-step instructions, traits particularly relevant for workflow synthesis. \emph{DeepSeek-V3} is selected for its strong performance in open-source benchmarks and its alignment with code-centric modeling, making it a competitive and transparent alternative to proprietary models. We evaluate each model using its default configuration settings, unless explicitly stated otherwise. This decision reflects the perspective of domain scientists, who may not possess expertise in adjusting model parameters. Requiring manual configuration could introduce unnecessary complexity and create additional barriers to practical use in real-world scientific settings.

We select ten bioinformatics workflows that represent a diversity of real-world data analysis tasks and domain-specific complexities (Table \ref{tab:sampleworkflows}). These workflows are sourced from two highly regarded and widely used community platforms: the \emph{Galaxy Training Network (GTN)} and the \emph{nf-core} initiative \cite{Hiltemann_2023, ewels2020nf}. GTN workflows offer clearly documented, step-by-step tutorials that cover core bioinformatics applications such as \emph{RNA-seq, variant calling, and quality control}. These workflows are ideal for evaluating how well LLMs can assist practitioners in reproducing structured analyses from natural language descriptions. In contrast, nf-core workflows are community-curated, production-ready pipelines built using Nextflow. They target high-throughput, reproducible research in environments that demand robustness, modularity, and scalability. Each nf-core pipeline adheres to strict guidelines, including automated testing, documentation standards, and best practices in computational reproducibility. Selecting workflows from nf-core allows us to evaluate LLM performance in replicating more sophisticated, script-based analyses essential for bioinformaticians.
By incorporating workflows from both GTN and nf-core, we ensure our evaluation captures a broad spectrum of user expertise, computational complexity, and bioinformatics domains. This dual-source approach enables a comprehensive assessment of LLMs' capabilities across graphical and code-based environments.

To address \textbf{RQ1}, we investigate the extent to which state-of-the-art LLMs can support the understanding and development of workflows in the bioinformatics domain. Our evaluation targets two essential capabilities: (1) \emph{foundational understanding of workflows and workflow systems}, and (2) \emph{construction of complete, domain-appropriate workflows}. To assess this, we begin by posing a set of fundamental questions (Table \ref{tab:fundamental_questions_of_SWS_and_SW}) that probe LLMs’ conceptual grasp of workflows, SWSs, and their core components, including tasks, tools, and data flow. This forms a baseline for determining whether models can meaningfully reason about workflows before generating them. We extend this analysis by evaluating each model's comprehension of \emph{platform-specific background knowledge} for our selected SWSs, using targeted questions (Table \ref{tab:sws_background_questions}) tailored to each platform's functionality, benefits, and architectural components. Additionally, we design \emph{workflow-specific background prompts} (Tables \ref{tab:galaxy_background_questions} and \ref{tab:nfcore_background_questions}) that correspond to representative bioinformatics workflows (Table~\ref{tab:sampleworkflows}). These background questions reflect the type of biological and computational knowledge a domain scientist needs to understand before workflow creation and are essential for evaluating the contextual readiness of LLMs to generate accurate workflows. This step served two purposes: (1) \emph{to confirm the LLM’s baseline understanding of the domain context}, and (2) \emph{to prime the model with relevant knowledge before executing the workflow prompt}. These background questions are curated based on domain expertise and established tutorials. While it is possible to pose many more questions to comprehensively assess LLM knowledge, our chosen set strikes a balance between breadth and depth. These questions are carefully curated to reflect key conceptual, platform-specific, and workflow-specific knowledge areas that a typical bioinformatics practitioner would need to grasp prior to workflow development. By focusing on essential topics, such as core workflow components, the principles of Galaxy and Nextflow, and the biological and computational context of each selected workflow, we ensure that the evaluation remains both focused and meaningful. Our goal is not to exhaustively test all possible knowledge domains, but to assess whether LLMs can understand and reason through the types of background information most relevant for supporting real-world workflow construction. This strategic sampling allows us to derive practical insights while keeping the evaluation tractable and grounded in realistic user scenarios.

\begin{table*}[htbp]
\vspace{-.5 em}
    \small
    \centering
    \caption{Overview of the used workflows. (W $\rightarrow$ Workflow)}
    \label{tab:sampleworkflows}
    \begin{tabularx}{\textwidth}{clXl}
    \toprule
        SWS & WID. & Workflow & Link \\
    \midrule
    Galaxy & W1 & Which coding exon has the highest number of single-nucleotide polymorphisms on human chromosome 22? & \cite{hiltemann2024galaxy} \\
           & W2 & How to get from peak regions to a list of gene names? & \cite{pajon2024peaks} \\
           & W3 & Do genes on opposite strands ever overlap? If so, how often? & \cite{clements2024introduction} \\
           & W4 & How can the quality of NGS raw data be assessed, what parameters should be checked, and how can the quality of the dataset be improved? & \cite{sequence-analysis-quality-control} \\
           & W5 & Which genes are on a draft bacterial genome, and which other genomic components can be found on a draft bacterial genome? & \cite{batut2024bacterial} \\
    \midrule
    Nextflow & W1 & nf-core/demo: This bioinformatics pipeline performs quality control and reporting on sequencing data. & \cite{christopher_hakkaart_2024_13951181} \\
    & W2 & nf-core/fetchngs: It is a bioinformatics pipeline to fetch metadata and raw FastQ files from databases. & \cite{harshil_patel_2024_10728509} \\
             & W3 & nf-core/methylseq: This workflow is used for Methylation (Bisulfite) sequencing data. & \cite{phil_ewels_2024_14502249} \\
             & W4 & nf-core/rnaseq: This workflow is used to analyse RNA sequencing data with a reference genome and annotation. & \cite{harshil_patel_2024_14537300} \\
             & W5 & nf-core/phyloplace: This workflow performs phylogenetic placement with EPA-NG. & \cite{daniel_lundin_2025_14906186} \\
    \bottomrule
    \end{tabularx}
\end{table*}

Following the comprehension phase, we evaluate the ability of each LLM to generate end-to-end workflows using three widely adopted prompting strategies (Details are in RQ3):(1) \textbf{Instruction-only}, which provides a direct task specification (e.g., Generate a Nextflow pipeline to analyze bisulfite sequencing data.); (2) \textbf{Role-based}, which frames the model as a domain expert (e.g., You are a bioinformatics workflow expert. Write a Galaxy workflow to determine gene overlap on opposite strands....); and (3) \textbf{Chain-of-thought}, which encourages step-by-step reasoning and decomposition of the task (e.g., Let’s break this task down step-by-step to fetch metadata and raw FastQ files....), which guides the model through step-by-step planning and justification. These strategies emulate realistic human-LLM interactions and provide insight into how prompting affects output quality and completeness. Each prompt is designed to simulate a domain scientist seeking assistance in workflow design. The evaluation process begins with an \emph{instruction-only} prompt. If the resulting workflow is valid, complete, and aligns with community baselines (e.g., GTN or nf-core \cite{hiltemann2023galaxy, ewels2020nf}), we do not proceed to more elaborate prompts. Otherwise, we escalate to role-based and CoT strategies to improve output quality. Across all tasks, we assess LLM-generated responses for \emph{syntactic correctness, executability, completeness, and domain relevance}.

In summary, our approach to RQ1 assesses whether LLMs can accurately understand the scientific context, correctly answer foundational and background questions, and construct usable workflows across a spectrum of bioinformatics tasks. Detailed results are described in Section \ref{results-of-prompts-to-pipeline}, including prompt-response summaries, workflow validation details, and model-specific performance insights.

To address \textbf{RQ2}, we systematically evaluate the workflows generated by each LLM along three core dimensions that define the practical utility of scientific workflows in bioinformatics: \emph{completeness, correctness, and usability}. Our objective is not merely to determine whether an LLM can produce a workflow, but to assess how well the generated workflows align with real-world scientific and technical standards. For this purpose, we utilize the workflows listed in Table \ref{tab:sampleworkflows}. Each task is benchmarked against a curated, community-reviewed, and publicly documented implementation from authoritative sources such as the GTN and nf-core \cite{Hiltemann_2023, ewels2020nf}. To simulate realistic usage scenarios, workflows are generated through conversational prompting using three strategies: \emph{instruction-only}, \emph{role-based}, and \emph{chain-of-thought}. Prompts are applied incrementally, starting with instruction-only and escalating to more structured prompting only when necessary, reflecting typical user behavior where minimal input is preferred.

Two authors, with over five and nine years of experience in scientific workflows and workflow systems, respectively, manually evaluate each generated workflow. Assessments are based on public documentation, workflow specifications, and verified execution examples. In addition to technical accuracy, we examine narrative clarity, such as explanatory comments and the rationale behind tool selection, which enhances usability, particularly for novice users.

Overall, our study is designed to go beyond superficial correctness by measuring how closely LLM-generated workflows approximate expert-designed standards in terms of reproducibility, executability, and scientific relevance. The results offer insights into the practical viability of using LLMs for real-world workflow development across diverse tasks and user expertise levels.

Understanding the prompting strategy for interacting with LLMs is essential for achieving accurate, relevant, and reliable responses. Existing research \cite{ekin2023prompt, chang2024efficient, marvin2023prompt} has shown that different prompting techniques, such as \emph{zero-shot, few-shot, role-based, and chain-of-thought}, lead to varying levels of performance across tasks. Without a deliberate prompting strategy, user queries may yield vague, incomplete, or even incorrect workflows, ultimately limiting the practical utility of LLMs and increasing inefficiency. Therefore, it is essential to evaluate which prompting strategies produce the most accurate, complete, and executable workflows. To address the \textbf{(RQ3)}, we investigate prompting methods grounded in recent literature on prompt engineering and LLM behavior [e.g., \cite{ekin2023prompt, arora2022ask, reynolds2021prompt, wei2022chain, zhou2022large, marvin2023prompt, lin2024write, liu2023pre, chang2024efficient}]. Our study surveys the theoretical foundations of these strategies and their prior success in domains involving procedural generation, planning, and code synthesis. We then experimentally evaluate different prompting techniques across diverse workflow scenarios, refining them through iterative testing. In Section \ref{results-of-prompts-to-pipeline}, we summarize our findings, providing actionable insights and best practices to help workflow developers optimize their prompting approaches for workflow generation. We provide a practical guideline to support workflow developers who may not possess expertise in prompt engineering or fine-tuning large language models. Recognizing that these users often lack formal training in AI or programming, our guideline is designed to be accessible and actionable, enabling them to effectively adopt our prompting approach for workflow development without requiring deep technical knowledge.

\section{Results} \label{results-of-prompts-to-pipeline}
In this section, we present the results of our evaluation of LLMs across three research questions. We assess their ability to understand fundamental workflow concepts, generate complete and correct scientific workflows, and respond effectively to various prompting strategies. Our findings are based on qualitative and execution-based analyses using workflows from Table \ref{tab:sampleworkflows}.

\subsection{RQ1: To what extent can LLMs assist in the development of scientific workflows?}
\textbf{Motivation: }Workflows in bioinformatics are essential for organizing, automating, and scaling complex data analyses. However, designing such workflows typically requires a combination of domain knowledge, technical proficiency, and iterative refinement skills that are often distributed across interdisciplinary teams. For many domain scientists, particularly those from non-computational backgrounds, understanding and constructing workflows using systems like Galaxy or Nextflow presents a significant challenge. SWSs involve abstract representations of tasks, parameter configurations, and tools/modules interactions that may be difficult to grasp without prior experience. LLMs have demonstrated strong capabilities in natural language understanding, code generation, and step-by-step reasoning. Recent advancements in LLMs present an opportunity to bridge this gap.

\textbf{Approach: }To address this research question, we implement a structured three-phase evaluation designed to capture both the conceptual understanding and practical construction capabilities of LLMs in scientific workflow development. The first phase focuses on assessing whether the models can accurately interpret and articulate foundational concepts related to scientific workflows and SWSs, which are essential for meaningful and context-aware workflow generation. We curate a set of nine fundamental prompts (P1–P9), listed in Table~\ref{tab:fundamental_questions_of_SWS_and_SW}, reflecting common conceptual inquiries that a domain scientist might pose when engaging with workflows.

\begin{table}[htbp]
\vspace{-.5em}
\scriptsize
\centering
    \caption{Prompts used to evaluate LLMs’ understanding of core concepts in workflows and workflow systems.}
    \label{tab:fundamental_questions_of_SWS_and_SW}
    \begin{tabular}{cl}
    \toprule
    \textbf{Prompt ID} & \textbf{Question (Prompt)} \\
    \midrule
    P1 & What is a scientific workflow? \\
    P2 & Why are scientific workflows used in research? \\
    P3 & What are the main steps in a typical scientific workflow? \\
    P4 & What is a Scientific Workflow System (SWS)? \\
    P5 & How does an SWS help automate data analysis? \\
    P6 & What is a task or process in a workflow? \\
    P7 & What is input and output data in the context of a workflow? \\
    P8 & What is a workflow tool, and what does it do? \\
    P9 & What is meant by data flow or data dependency in workflows? \\
    \bottomrule
\end{tabular}
\vspace{-1em}
\end{table}

To further ensure contextual grounding, we supplement this phase with a second set of platform-specific background questions targeting Galaxy and Nextflow (Table~\ref{tab:sws_background_questions}). 
\begin{table}[htbp]
\scriptsize
\centering
\caption{Prompts used to assess LLMs’ understanding of Galaxy and Nextflow Workflow Systems.}
\label{tab:sws_background_questions}
\begin{tabular}{p{1cm}p{5.5cm}p{5.5cm}}
\toprule
\textbf{Prompt ID} & \textbf{Galaxy (Prompt)} & \textbf{Nextflow (Prompt)} \\
\midrule
P1 & What is the Galaxy platform, and how is it used in bioinformatics? & What is the Nextflow workflow system, and how is it applied in bioinformatics? \\
P2 & How does Galaxy support reproducibility and transparency in workflows? & How does Nextflow enable reproducibility and portability in bioinformatics pipelines? \\
P3 & What types of analyses can be performed using Galaxy (e.g., RNA-seq, variant calling)? & What types of bioinformatics analyses are commonly implemented using Nextflow (e.g., RNA-seq, metagenomics)? \\
P4 & What are the benefits of Galaxy for bioinformatics workflow? & What are the benefits of using Nextflow for building and scaling bioinformatics workflows? \\
P5 & What are the main components of the Galaxy Scientific Workflow System? & What are Process, Channel, Module, and Operator in Nextflow? \\
P6 & What is Galaxy ToolShed? & What is nf-core? \\
\bottomrule
\end{tabular}
\vspace{-1em}
\end{table}

These are intended to evaluate each LLM’s familiarity with the architecture, features, and design paradigms of the two selected SWSs. Importantly, these questions are submitted to the models prior to issuing any workflow generation prompts, thereby allowing the LLMs to align their reasoning with our experimental intent.
All background questions are posed to \emph{GPT-4o, Gemini 2.5 Flash, and DeepSeek-V3}. Each model is instructed to provide concise yet informative answers. The responses are independently reviewed by two authors with substantial domain expertise (over five and nine years of experience in scientific workflows, respectively) and are cross-referenced against authoritative literature and official documentation to verify their accuracy and relevance. The complete set of responses is made available in the supplementary materials \cite{alam_2025_16416384}.
\begin{table}[htbp]
\scriptsize
\centering
\caption{Background questions designed for Galaxy workflows to assess LLMs' foundational understanding before workflow generation. Each set corresponds to a specific use case.}
\label{tab:galaxy_background_questions}
\begin{tabular}{p{0.7cm}p{1cm}p{11cm}}
\toprule
\textbf{WID} & \textbf{Prompt ID} & \textbf{Background Question} \\
\midrule
W1 & P1 & What are nucleotides, chromosomes, exons, and SNPs? \\
   & P2 & What is a sequencing read, and how is its quality measured? \\
   & P3 & What is the difference between raw and processed sequencing data? \\
   & P4 & What is RNA splicing?\\
   & P5 & What is the difference between Exon and Intron?\\
\midrule
W2 & P1 & What is ChIP-seq, and what biological information does it provide? \\
   & P2 & What is a peak in genomic data, and how is it represented in a BED file? \\
   & P3 & What is gene annotation, and how is it used to associate peaks with genes? \\
   & P4 & What is the transcription start site (TSS), and why is it important in gene regulation? \\
   & P5 & What are the key file formats in peak annotation, and how do they differ? \\
   & P6 & What is the promoter region of genes? \\
   & P7 & Why is it important to identify target genes from genomic regions in research? \\
\midrule
W3 & P1 & What is the structure of double-stranded DNA, and what are the forward and reverse strands? \\
   & P2 & What are genome and chromosome? \\
   & P3 & What are reference genome and GENCODE? \\
   & P4 & How is strand orientation recorded in genome annotation files? \\
   & P5 & How do forward/reverse strands affect read mapping and quantification? \\
   & P6 & What does it mean for genes on opposite strands to overlap, and why is this an interesting biological question? \\
   & P7 & What are genomic intervals, and how is the overlap between features defined? \\
   & P8 & How can understanding gene strand orientation impact the interpretation of genetic information? \\
   & P9 & What’s the difference between sequence and annotation? \\
   & P10 & What are FASTA, BED, GTF, GFF3, and VCF? \\
   & P11 & What are GRCh37, GRCh38, hg19, and hg38? \\
\midrule
W4 & P1 & What is Sanger sequencing? Provide some examples. \\
   & P2 & What are primers in sequencing? \\
   & P3 & What is the AB1 sequence file? \\
   & P4 & What are CHD8 and AOPEP? \\
   & P5 & What are sense, antisense, and consensus sequences? \\
   & P6 & What is the reverse-complement sequence, and why compute it for the antisense strand? \\
   & P7 & Why is converting from AB1 sequence file to FASTQ essential? \\
\midrule
W5 & P1 & What is genome annotation, and why is it critical for bacterial genomics? \\
   & P2 & What is Bakta for bacterial genome annotation? \\
   & P3 & What are coding sequences (CDS), and what roles do rRNAs and tRNAs play? \\
   & P4 & What are plasmids? \\
   & P5 & What are integrons? \\
   & P6 & What are GFF3 and GenBank formats, and how are they used in annotation pipelines? \\
   & P7 & How do you evaluate the quality of an annotated genome? \\
\bottomrule
\end{tabular}
\vspace{-1em}
\end{table}
In the second phase of our evaluation, we focus on assessing LLMs' contextual understanding of specific bioinformatics workflows before prompting them to generate complete pipeline implementations. To formulate relevant background questions, we first implement each of the selected workflows by meticulously following the step-by-step instructions provided in the GTN and nf-core documentation. These tutorials are typically enriched with biological and computational context, which we carefully analyze to extract the foundational concepts required for informed workflow generation.
This preliminary implementation process serves two purposes: (1) \emph{it ensures that we, as evaluators, develop a deep understanding of each workflow's intent and computational logic}; and (2) \emph{it allows us to derive meaningful, targeted questions that reflect the type of background knowledge a domain scientist would need prior to developing a workflow}. The two authors independently review the documentation and execute each workflow, and then collaboratively synthesize key contextual elements into background prompts designed to test the LLMs' readiness for the task. The complete sets of background questions for workflows based on Galaxy and Nextflow are provided in Table~\ref{tab:galaxy_background_questions} and Table~\ref{tab:nfcore_background_questions}, respectively.

\begin{table}[htbp]
\vspace{-0.5em}
\centering
\scriptsize
\caption{Background questions designed for nf-core workflows to assess LLMs' foundational understanding before workflow generation. Each set corresponds to a specific pipeline.}
\label{tab:nfcore_background_questions}
\begin{tabular}{p{0.7cm}p{1cm}p{11cm}}
\toprule
\textbf{WID} & \textbf{Prompt ID} & \textbf{Background Question} \\
\midrule
W1 & P1 & What is the purpose of performing quality control on sequencing data? \\
   & P2 & What are sequencing adapters, and why is it necessary to remove them from raw reads? \\
   & P3 & How does read quality affect downstream bioinformatics analyses such as alignment or quantification? \\
   & P4 & What is the difference between single-end and paired-end reads, and how does it influence preprocessing?\\
   & P5 & Why is it important to visualize QC metrics before proceeding with downstream analysis? \\
   & P6 & What is FastQC, and what metrics does it generate for assessing sequencing read quality? \\
\midrule
W2 & P1 & What is NGS? \\
   & P2 & What are SRA, ENA, DDBJ, and GEO and why are they important for NGS data retrieval? \\
   & P3 & What is a FASTQ file, and what type of biological data does it contain? \\
\midrule
W3 & P1 & What is DNA methylation and why is bisulfite sequencing used to measure it? \\
   & P2 & What is genome alignment? \\
   & P3 & What are the challenges in aligning bisulfite-converted reads to a reference genome? \\
   & P4 & What types of input data (e.g., single-end or paired-end FASTQ) are required for methylation analysis? \\
   & P5 & How are methylation levels quantified and visualized across genomic regions? \\
   & P6 & Briefly describe Bismark and BWA-meth? \\
\midrule
W4 & P1 & What is RNA-seq and how does it help quantify gene expression? \\
   & P2 & What is a reference genome and annotation? \\
   & P3 & What is a gene expression matrix? \\
   & P4 & What preprocessing steps (e.g., adapter trimming, QC) are necessary before alignment? \\
   & P5 & What are transcript-level vs. gene-level quantifications and when are they used? \\
\midrule
W5 & P1 & What is phylogenetic placement, and how does it differ from reconstructing a full phylogenetic tree? \\
   & P2 & Why is phylogenetic placement useful for short or fragmentary sequences? \\
   & P3 & What is EPA‑NG? \\
   & P4 & What are query and reference sequences? \\
\bottomrule
\end{tabular}
\vspace{-1em}
\end{table}

In the final phase of our study addressing RQ1, we task the LLMs with generating workflows using Galaxy or Nextflow, depending on the platform associated with each selected use case. This stage is designed to evaluate the extent to which LLMs can synthesize and operationalize workflow knowledge into executable pipelines. Beyond verifying syntactic validity, we focus on whether the generated workflows exhibit appropriate tool selection, logical task sequencing, and adherence to community-established conventions for reproducibility and usability.
\begin{table}[htbp]
\vspace{-1em}
\scriptsize
\caption{Prompts used for Galaxy workflow generation across different use cases.}
\label{tab:galaxy_prompts}
\centering
\begin{tabularx}{\columnwidth}{p{0.12\columnwidth}X}
\toprule
\textbf{Workflow ID} & \textbf{Prompt Type and Content} \\
\midrule

\textbf{W1} & 
\textbf{Instruction-Only:} Create a Galaxy workflow that identifies the coding exon with the highest number of single-nucleotide polymorphisms (SNPs) on human chromosome 22. Include all necessary tools, input data types, and expected outputs. \\
\midrule

\textbf{W2} & 
\textbf{Instruction-Only:} Create a Galaxy workflow to identify target genes from ChIP-seq peak regions. Use the dataset \texttt{GSE37268\_mof3.out.hpeak.txt.gz} and a mouse gene annotation file from UCSC. 

\textbf{Role-Based:} You are a bioinformatics workflow developer. Create a Galaxy workflow to identify target genes from ChIP-seq peak regions. The peak file contains numeric chromosome identifiers (e.g., 1–20 for X, 21 for Y) instead of UCSC-style labels. Preprocess the chromosome field accordingly to ensure compatibility with standard annotations. \\
\midrule

\textbf{W3} & 
\textbf{Instruction-Only:} Create a Galaxy workflow to determine whether genes on opposite DNA strands overlap, using a gene annotation dataset in BED format. \\
\midrule

\textbf{W4} & 
\textbf{Instruction-Only:} Create a Galaxy workflow to perform quality control on raw next-generation sequencing (NGS) data, ensuring the dataset is assessed and improved for high-quality downstream analysis. \\
\midrule

\textbf{W5} & 
\textbf{Instruction-Only:} Create a Galaxy workflow to annotate an assembled bacterial genome provided in FASTA format. Use tools to predict genes and functional elements, identify plasmid replicons, detect integron structures, and find insertion sequences. Convert tool-specific outputs into GFF3 format and load all annotations into a genome browser for interactive visualization.  

\textbf{Role-Based:} You are a bioinformatics workflow developer working within the Galaxy platform. Create a workflow to annotate a draft bacterial genome, evaluate the annotation, format the outputs for visualization, and visualize the results.  

\textbf{Chain-of-Thought:} We have a draft bacterial genome in FASTA format. First, identify genomic components such as CDS, rRNAs, and tRNAs with structural and functional annotations. Save outputs in standard formats for downstream use (e.g., feature tables, protein/nucleotide sequences). Check annotation completeness and summarize statistics. Finally, organize outputs for compatibility with genome browsers. Based on this reasoning, generate a Galaxy workflow that completes these tasks. \\
\bottomrule
\end{tabularx}
\vspace{-0.1em}
\end{table}
To systematically guide this generation process, we adopt a three-tiered prompting framework that varies in the degree of contextual scaffolding: \emph{instruction-only}, \emph{role-based}, and \emph{chain-of-thought (CoT)} prompting. The instruction-only strategy presents a concise directive without any added context, aiming to simulate a general user query. The role-based approach enhances this by assigning the LLM a specific professional identity, typically that of a bioinformatics expert, to encourage more domain-informed reasoning. Finally, the CoT prompting strategy explicitly encourages step-by-step workflow planning, justification of tool/module selection, description of input/output formats, and commentary on task dependencies and execution logic.

\begin{table}[htbp]
\vspace{-1em}
\centering
\scriptsize
\caption{Prompts used for Nextflow workflow generation across different use cases}
\label{tab:nextflow_prompts}
\begin{tabularx}{\columnwidth}{p{0.12\columnwidth}X}
\hline
\textbf{Workflow ID} & \textbf{Prompt Type and Content} \\
\hline
\textbf{W1} & 
\textbf{Instruction-Only:} We want to create a workflow using Nextflow that performs quality assessment of FASTQ files, preprocesses sequencing data, and generates an aggregated summary of the results. The input data should be provided via the sample sheet available at \href{https://raw.githubusercontent.com/nf-core/test-datasets/viralrecon/samplesheet/samplesheet\_test\_illumina\_amplicon.csv}{samplesheet\_test\_illumina\_amplicon.csv}. Please outline the step-by-step modules required to build this workflow. \\
\hline
\textbf{W2} & 
\textbf{Instruction-Only:} We want to create a Nextflow workflow that takes a CSV file containing sample database IDs (e.g., SRA, ENA, GEO, or DDBJ) as input, retrieves associated metadata, and downloads raw FASTQ files. Please outline the step-by-step modules to create the workflow. \\
\hline
\textbf{W3} & 
\textbf{Instruction-Only:} We want to create a Nextflow workflow for DNA methylation analysis using bisulfite-converted sequencing (BS-seq) data. The workflow should begin by pre-processing raw reads provided in a CSV file containing FASTQ input information, followed by alignment to a reference genome and comprehensive quality control assessments. Could you please provide a step-by-step outline of the modules required to construct this workflow?

\textbf{Role-Based:} You are a bioinformatics workflow developer with expertise in building reproducible pipelines using Nextflow and nf-core modules. Your task is to design a Nextflow workflow for DNA methylation analysis using bisulfite-converted sequencing (BS-seq) data. The workflow should begin by pre-processing raw reads specified in a CSV file (containing sample IDs and FASTQ paths), then proceed to align reads to a reference genome, extract methylation information, and generate comprehensive quality control reports. Please provide a step-by-step outline of the modules required to build this workflow, using only officially available nf-core or Nextflow modules, and include their exact module names where appropriate. \\
\hline
\textbf{W4} & 
\textbf{Instruction-Only:} We want to create a Nextflow workflow for RNA-seq analysis that performs quality control, adapter trimming, alignment, quantification, and differential expression analysis. The workflow should accept a CSV sample sheet containing FASTQ file links and genome references as input. It must include quality checks, reference genome indexing if needed, and gene/transcript quantification, and produce a gene expression matrix and extensive QC report using appropriate tools. Please provide a step-by-step outline of the modules to construct this workflow. \\
\hline

\textbf{W5} & 
\textbf{Instruction-Only:} We want to create a Nextflow workflow for the phylogenetic placement of query sequences onto a reference tree, supporting two operational modes. In placement-only mode, the workflow uses a query FASTA file, a reference alignment, a rooted phylogenetic tree in Newick format, an evolutionary model, and optionally a taxonomy file for classification. In search-plus-placement mode, it begins with a large unfiltered FASTA file and HMM profiles to identify candidate sequences, followed by alignment, placement, classification, and reporting using the same reference inputs. The workflow aligns selected sequences, places them onto the tree, classifies their positions, and generates summary reports and visualizations. Could you please provide a step-by-step outline of the modules required to construct this workflow? \\
\hline
\end{tabularx}
\vspace{-0.5em}
\end{table}

For instance, in the Galaxy-based workflow \emph{W2: Peaks to Genes}, the CoT prompt elicits structured responses in which the LLM summarizes the biological objective, selects appropriate annotation tools (e.g., ChIPseeker), and provides justifications for each component. In the case of Nextflow, the \emph{W4: nf-core/rnaseq} workflow, the CoT strategy effectively prompts the LLM to outline the modular design, describe input channels, and specify configuration parameters aligned with Nextflow syntax.

We apply this prompting strategy uniformly across all considered workflows and LLMs, ensuring methodological consistency for fair comparison. This design allows us to assess how varying degrees of prompt specificity influence the LLMs’ ability to generate accurate, complete, and usable workflows in Galaxy and Nextflow. Ultimately, this phase serves as a bridge between conceptual understanding and practical construction, offering insight into the generative capabilities of LLMs within real-world bioinformatics contexts.

Drawing from existing literature on effective prompting strategies (discussed in RQ3), we design and apply structured prompts to guide LLMs in workflow generation. Tables~\ref{tab:galaxy_prompts} and~\ref{tab:nextflow_prompts} present representative prompts used to guide the LLMs in generating our workflows. These examples illustrate how varying levels of contextual scaffolding, ranging from instruction-only to role-based and chain-of-thought prompts, are employed to elicit structured and contextually appropriate responses. While the prompts shown are specific to selected case studies, a consistent prompting strategy is applied across all workflows evaluated in this study, as detailed in the results section. The complete set of prompts and corresponding LLM outputs are included in the supplementary material to promote transparency and support reproducibility.

To validate the generated workflows, we utilize the Galaxy web interface available at \url{https://usegalaxy.org/}, which offers a fully hosted environment for executing and testing Galaxy-based workflows. For Nextflow workflows, we follow the official installation procedures provided in the documentation at \href{https://www.nextflow.io/docs/latest/install.html}{nextflow.io/docs/latest/install.html}. Additionally, for users seeking a lightweight or cloud-based setup, workflows can be executed through GitHub Codespaces using the pre-configured Nextflow training environment available at \href{https://github.com/codespaces/new/nextflow-io/training?quickstart=1&ref=2.2}{github.com/codespaces/nextflow-io/training}. This setup offers a streamlined alternative to local installation and ensures reproducibility in a controlled execution environment.

We adopt a stepwise prompting strategy designed to minimize unnecessary complexity while ensuring the generation of accurate and complete workflows. Each workflow generation task begins with an instruction-only prompt. If the output from this initial prompt aligns with the expected workflow structure, based on baselines such as GTN or nf-core documentation, we do not proceed with further prompting. However, if the workflow is incomplete, incorrect, or misaligned with the baseline, we escalate to a role-based prompt to encourage more domain-specific reasoning. If this also fails to yield a satisfactory result, we employ a chain-of-thought (CoT) prompt, which guides the model through a step-by-step reasoning process. This tiered prompting strategy enables a principled trade-off between prompting complexity and output quality, while also allowing us to assess the minimal level of guidance required for successful workflow generation.

\textbf{Results of RQ1: }

We present a set of core conceptual prompts to evaluate the LLMs' foundational understanding of workflows and workflow systems (Table~\ref{tab:fundamental_questions_of_SWS_and_SW}). All three models generate technically accurate responses and are aligned with each question's intended meaning. To further assess their effectiveness, we (two authors, with over five and nine years of experience) evaluate the responses using six qualitative criteria: \emph{correctness, clarity, completeness, terminology, structure, and usability}. A comparison of their performance is summarized in Table~\ref{tab:llm_comparison_fundamental_concepts}. 

While all models satisfied the correctness and terminology criteria, notable differences emerged across the remaining dimensions. GPT-4o demonstrated strong clarity and accessibility, offering responses that are well-suited for foundational learning and novice users. Gemini 2.5 Flash provided the most comprehensive and academically precise answers, making it ideal for users with intermediate experience or those seeking formal educational material. DeepSeek-V3 delivered contextually accurate responses enriched with practical examples; however, its outputs are occasionally verbose and informally structured, limiting their immediate instructional usability without refinement.

We next evaluate the responses to the platform-specific background prompts designed to assess LLMs’ contextual understanding of Galaxy and Nextflow (Table~\ref{tab:sws_background_questions}). The evaluation results, summarized in Table~\ref{tab:llm_comparison_about_sws}, reveal distinct strengths and limitations across the three models.
GPT-4o produces accurate, well-structured, and beginner-friendly responses characterized by clear language and logical organization. However, it occasionally omits advanced implementation details such as containerization strategies and execution environments, which are important for reproducibility and deployment. Gemini 2.5 Flash delivers the most academically rigorous and comprehensive responses, offering precise explanations that cover tool integration, ecosystem capabilities, and platform-specific components. Its slightly formal tone and structured presentation make it particularly well-suited for intermediate users or instructional contexts. DeepSeek-V3 performs well in technical accuracy and real-world applicability, often referencing specific tools and platforms such as ToolShed and WorkflowHub. However, its responses are frequently verbose, rely heavily on bullet-point formatting, and lack the narrative flow necessary for effective instructional use. Overall, Gemini provides the most complete and balanced output, GPT-4o excels in clarity and accessibility for novice users, and DeepSeek-V3 offers valuable practical insight but would benefit from refinement to improve instructional usability.
\newcommand{\greencheck}{\textcolor{green!60!black}{\faCheck}}
\newcommand{\yellowwarn}{\textcolor{orange!80!black}{\faExclamationTriangle}}

\begin{table*}[htbp]
\vspace{-.5em}
\scriptsize
\centering
\caption{Comparison of LLM Responses across Evaluation Criteria for Fundamental Concepts (Table \ref{tab:fundamental_questions_of_SWS_and_SW} )}
\begin{tabular}{p{2.2cm}p{3.9cm}p{3.9cm}p{4.2cm}}
\toprule
\textbf{Criterion} & \textbf{GPT-4o} & \textbf{Gemini 2.5 Flash} & \textbf{DeepSeek-V3} \\
\midrule
\textbf{Correctness} & \greencheck\  Accurate but basic definitions & \greencheck\ Most precise with academic rigor & \greencheck\ Technically sound with practical examples \\
\midrule
\textbf{Clarity} & \greencheck\ Very clear and straightforward & \greencheck\ Clear but occasionally formal & \yellowwarn\ Clear but Verbose \\
\midrule
\textbf{Completeness} & \yellowwarn\ Covers basics but lacks depth i.e., omits advanced concepts like validation & \greencheck\ Comprehensive answers with good coverage of key concepts & \greencheck\ Covers most concepts explicitly, often using enumeration \\
\midrule
\textbf{Terminology} & \greencheck\ Uses familiar and readable technical terms. & \greencheck\ Uses formal and precise academic terminology & \greencheck\ Appropriate with real-world SWSs (e.g., Galaxy, KNIME) \\
\midrule
\textbf{Structure} & \greencheck\ Consistent and clean & \greencheck\ Well-organized & \yellowwarn\ Often uses lists and informal bullets \\
\midrule
\textbf{Usability} & \greencheck\ Excellent for beginner tutorials & \greencheck\  Good for intermediate audiences & \yellowwarn\ Needs cleanup for instructional material \\
\bottomrule
\end{tabular}
\label{tab:llm_comparison_fundamental_concepts}
\vspace{-1.0em}
\end{table*}
To evaluate the ability of LLMs to address workflow-specific background questions, we provide a set of foundational biological and technical prompts (Table~\ref{tab:galaxy_background_questions}) related to representative Galaxy workflows. 
\begin{table*}[htbp]
\vspace{-0.2em}
\centering
\scriptsize
\caption{Comparison of LLM Responses across Evaluation Criteria for Background Questions about Galaxy and Nextflow (Table \ref{tab:sws_background_questions})}
\begin{tabular}{p{2.2cm}p{3.9cm}p{3.9cm}p{4.2cm}}
\toprule
\textbf{Criterion} & \textbf{GPT-4o} & \textbf{Gemini 2.5 Flash} & \textbf{DeepSeek-V3} \\
\midrule
\textbf{Correctness} & \greencheck\ Accurate but concise; omits container/cloud details & \greencheck\ Most precise with deep tool/system understanding & \greencheck\ Correct with real examples from SWSs \\
\midrule
\textbf{Clarity} & \greencheck\ Clear and beginner-friendly & \greencheck\ Clear with slight formal tone & \yellowwarn\ Verbose and uses list-heavy structure \\
\midrule
\textbf{Completeness} & \yellowwarn\ Covers basics; lacks advanced implementation details & \greencheck\ Thorough with ecosystem, portability, and usage depth & \greencheck\ Covers broad scope and tool support explicitly \\
\midrule
\textbf{Terminology} & \greencheck\ Accessible language for broad users & \greencheck\ Formal academic terminology well-used & \greencheck\ Includes domain-specific terms and tools (e.g., ToolShed, WorkflowHub) \\
\midrule
\textbf{Structure} & \greencheck\ Narrative, well-flowing format & \greencheck\ Well-structured with natural coherence & \yellowwarn\ Checklists over narrative, uneven depth \\
\midrule
\textbf{Usability} & \greencheck\ Suitable for introductory tutorials & \greencheck\ Great for intermediate bioinformatics learners & \yellowwarn\ Needs formatting edits for teaching or documentation \\
\bottomrule
\end{tabular}
\label{tab:llm_comparison_about_sws}
\vspace{-1.5em}
\end{table*}
The results of this evaluation (Table~\ref{tab:llm_comparison_galaxy}) indicate that Gemini 2.5 Flash consistently provides the most accurate and comprehensive responses. For example, in response to \emph{W2–P3 (What is gene annotation, and how is it used to associate peaks with genes?)}, Gemini offers a detailed explanation encompassing both functional annotation and genomic coordinate mapping, whereas GPT-4o and DeepSeek-V3 provide less insight into the mechanisms linking peaks to nearby genes. Similarly, for W5–P2 (\emph{What is Bakta for bacterial genome annotation?}), Gemini elaborates on database usage and output formats, while other models only describe its general purpose.

GPT-4o, although less comprehensive, stands out for its clarity and ease of understanding, making it particularly suitable for beginners. For example, its explanation of W1–P1 (\emph{What are nucleotides, chromosomes, exons, and SNPs?}) was concise, well-structured, and accessible to users with minimal background in genomics. On the other hand, DeepSeek-V3, while generally accurate, shows inconsistencies in depth and completeness. In W3–P6 (\emph{What does it mean for genes on opposite strands to overlap?}), DeepSeek fails to address key implications of antisense transcription and regulatory complexity that are captured by Gemini. These findings indicate that Gemini 2.5 Flash is best suited for users seeking technical rigor, GPT-4o excels in communication for broader audiences, and DeepSeek-V3 may require improvement in consistency and contextual depth for bioinformatics education and support tasks.
\begin{table*}[htbp]
\scriptsize
\vspace{-0.5em}
\centering
\caption{Comparison of LLM Responses for Background Prompts in Galaxy Workflows}
\begin{tabular}{p{2.3cm}p{3.9cm}p{3.9cm}p{4.2cm}}
\toprule
\textbf{Criterion} & \textbf{GPT-4o} & \textbf{Gemini 2.5 Flash} & \textbf{DeepSeek-V3} \\
\midrule
\textbf{Correctness} & \greencheck\ Generally accurate; some minor conflation in technical distinctions & \greencheck\ Highly accurate and aligned with academic definitions & \greencheck\ Mostly correct but occasionally vague or incomplete \\
\midrule
\textbf{Clarity} & \greencheck\ Very clear and accessible to beginners & \yellowwarn\ Clear but often formal or verbose & \yellowwarn\ Mixed clarity; some explanations too shallow or ambiguous \\
\midrule
\textbf{Completeness} & \yellowwarn\ Basic coverage; misses deeper insights & \greencheck\ Comprehensive, includes biological relevance and technical context & \yellowwarn\ Omits some intermediate concepts or format distinctions \\
\midrule
\textbf{Conciseness} & \greencheck\ Concise and to the point & \yellowwarn\ Informative but often lengthy & \greencheck\ Generally concise but may lack depth \\
\midrule
\textbf{Terminology} & \greencheck\ Uses common and understandable terms & \greencheck\ Uses precise and domain-specific vocabulary & \yellowwarn\ Uses general terms; sometimes lacks specificity \\
\midrule
\textbf{Mistakes/Errors} & Minor blending of concepts in definitions & Occasionally overuses jargon without explanation & Misses important context (e.g., gene–peak associations) \\
\midrule
\textbf{Overall Verdict} & \textbf{Most user-friendly} & \textbf{Best academic rigor and depth} & \textbf{Needs improvement in consistency and detail} \\
\bottomrule
\end{tabular}
\label{tab:llm_comparison_galaxy}
\vspace{-1em}
\end{table*}

To assess how well LLMs explain workflow-specific background concepts in Nextflow pipelines, we analyze their responses using representative Nextflow workflows. These prompts covered foundational biological and computational topics relevant to our workflows. Table~\ref{tab:llm_comparison_nextflow} summarizes the comparative performance across. 
\begin{table*}[htbp]
\vspace{-.5em}
\scriptsize
\centering
\caption{Comparison of LLM Responses for Background Prompts in Nextflow Workflows}
\begin{tabular}{p{2.3cm}p{3.9cm}p{3.9cm}p{4.2cm}}
\toprule
\textbf{Criterion} & \textbf{GPT-4o} & \textbf{Gemini 2.5 Flash} & \textbf{DeepSeek-V3} \\
\midrule
\textbf{Correctness} & \greencheck\ Generally correct with some minor oversights in tool-specific context & \greencheck\ Highly accurate and comprehensive across workflows & \greencheck\ Mostly accurate with detailed examples \\
\midrule
\textbf{Clarity} & \greencheck\ Clear and easy to follow for most users & \yellowwarn\ Slightly formal tone but readable & \yellowwarn\ Occasionally dense and list-heavy \\
\midrule
\textbf{Completeness} & \yellowwarn\ Omits details like parameter flags and metadata resolution & \greencheck\ Covers technical tools, configurations, and context thoroughly & \yellowwarn\ Strong examples, but some answers are tool-biased or partial \\
\midrule
\textbf{Terminology} & \greencheck\ Balanced use of domain-relevant terms & \greencheck\ Advanced bioinformatics vocabulary used appropriately & \yellowwarn\ Inconsistent: sometimes general, sometimes overly specific \\
\midrule
\textbf{Conciseness} & \greencheck\ Well-balanced for instructional use & \yellowwarn\ Verbose in longer answers & \greencheck\ Compact but requires interpretation \\
\midrule
\textbf{Mistakes/Errors} & Minor: occasionally omits integration aspects or setup steps & Few: may overwhelm novices with detail & Minor: tool links are correct but lack rationale \\
\midrule
\textbf{Overall Verdict} & \textbf{Best for clarity and beginners} & \textbf{Best for completeness and experts} & \textbf{Good for technical detail, less suited for guided explanation} \\
\bottomrule
\end{tabular}
\label{tab:llm_comparison_nextflow}
\vspace{-1.5em}
\end{table*}

Gemini 2.5 Flash addresses complex topics such as metadata retrieval in \emph{fetchngs}, reference genome handling in \emph{phyloplace}, and bisulfite alignment nuances in \emph{methylseq}, often incorporating tool-specific details, flags, and file structure references. For example, Gemini correctly explained the importance of ENA API calls for resolving sample metadata and provided detailed descriptions of output artifacts. Its responses are comprehensive, though sometimes overly verbose or formal, which could be challenging for novice users. GPT-4o, by contrast, prioritizes clarity and accessibility. It provides succinct, beginner-friendly explanations that are logically structured and easy to follow. For example, its explanation of the FASTQ format and the rationale for adapter trimming in RNA-seq workflows is precise and instructional. However, GPT-4o occasionally lacks depth in areas such as the resolution of experiment-level identifiers or the rationale behind multiple HMM profiles in phyloplace. DeepSeek-V3 delivered technically correct responses with a strong practical orientation. It excels in referencing command-line tools, directory structures, and usage examples. In the context of fetchngs, for instance, it correctly described MD5 checksum validation and discussed the role of sample mapping fields. However, its answers often leaned toward being list-heavy and missed some of the conceptual explanations, such as the biological rationale behind bin refinement in MAG analysis or how GAPPA supports phylogenetic visualization. 

In summary, while all three LLMs perform reasonably well in covering core concepts, Gemini 2.5 Flash is best suited for users seeking comprehensive, GPT-4o excels in readability and instructional clarity, and DeepSeek-V3 offers detailed practical insights but requires refinement for conceptual completeness and consistency.

\subsubsection{Workflow Development Using LLMs for Galaxy: }
For Workflow W1: \emph{Identifying the exon with the highest number of SNPs on human chromosome 22} within the Galaxy SWS, we first manually construct and execute the workflow using the Galaxy platform. This was done by following the step-by-step guidance provided in the Galaxy Training Network (GTN) tutorial \cite{hiltemann2024galaxy}. Utilizing the sample dataset provided by GTN, we determine that the exon \emph{ENST00000253255.7\_cds\_0\_0\_chr22\_46256561\_r} contains the highest number of SNPs (27 in total) on human chromosome 22. Subsequently, we apply the instruction-only prompt (W1, Table~\ref{tab:galaxy_prompts}) to GPT-4o, asking it to generate the corresponding workflow steps. The model successfully produces a logically ordered Galaxy workflow that correctly identifies the same exon. Notably, while the GTN tutorial assumes that the input files are already restricted to chromosome 22, GPT-4o explicitly includes filtering steps to isolate chromosome 22 from genome-wide datasets. This reflects a more realistic real-world scenario and demonstrates the model’s capacity to generalize beyond the assumptions of curated tutorials. In terms of tool selection, we observe minor differences in the implementation details. The GTN workflow uses the \emph{bedtools intersect intervals} tool to compute overlaps between SNPs and exons, whereas GPT-4o suggests the use of the \emph{Join} tool to perform the same operation. Additionally, while the GTN workflow employs the \emph{Datamash} tool to count SNPs per exon, GPT-4o proposes using the \emph{Group} tool. Despite these differences in tool choices, the overall logic and structure of the workflow remain correct and effective.

To evaluate the performance of Gemini 2.5 Flash, we use the same instruction-only prompt used previously. The model generates a clear and well-structured workflow, offering step-by-step guidance that includes both data acquisition and filtering procedures. Similar to the GTN tutorial, Gemini 2.5 Flash recommends using the \emph{bedtools intersect intervals} tool to compute overlaps between SNPs and exons. However, for aggregating SNP counts per exon, it opted for the \emph{Group} tool instead of the \emph{Datamash} tool used in GTN. The workflow also includes the use of the \emph{Sort} tool, a common step across all three implementations: GTN, GPT-4o, and Gemini 2.5 Flash. Upon executing the workflow generated by Gemini 2.5 Flash using the same GTN dataset, we obtain the same correct result, identifying the exon \emph{ENST00000253255.7\_cds\_0\_0\_chr22\_46256561\_r} with 27 SNPs, as observed in the baseline and GPT-4o workflows.

For DeepSeek-V3, we again apply the same instruction-only prompt and are successfully able to generate the steps to develop correct and executable Galaxy workflow. Similar to GPT-4o, DeepSeek-V3 recommends the use of the \emph{Intersect} tool to compute overlaps between SNPs and exon regions. Additionally, the model offers flexibility by suggesting alternative tools for certain tasks. For example, to count the number of SNPs per exon, it proposes using either the \emph{Group} or \emph{Count} tool, both valid and appropriate choices. For identifying the exon with the highest SNP count, it recommends the combination of \emph{Sort} followed by \emph{Select}, which aligns with standard Galaxy practices. Notably, DeepSeek-V3 also provides supplementary information beyond the core task. It includes guidance on how to export, import, and execute the workflow within the Galaxy platform, along with suggestions for potential public data sources for exons and SNPs. While such details go beyond the minimal workflow requirements, they enhance usability for novice users. Overall, DeepSeek-V3 produces a functionally correct workflow from the instruction-only prompt, demonstrating its capability to support bioinformatics workflow design.

Given that the instruction-only prompt produce valid and complete workflows across all three LLMs, we do not proceed with the more elaborate Role-based or CoT prompting strategies for this task. This outcome indicates that for well-defined and well-scoped problems like Workflow W1 (Galaxy), minimal prompting is sufficient to guide advanced LLMs in generating accurate and executable workflows.

\emph{While all three LLMs, successfully generate correct and executable workflows from the instruction-only prompt, Gemini 2.5 Flash stands out for its combination of technical accuracy and instructional clarity. Its response closely aligns with the GTN tutorial, uses appropriate tools, and provides a structured, step-by-step guide that balances detail with usability. Although GPT-4o produces a concise and accurate workflow with thoughtful generalizations, such as including chromosome filtering, it lacks some of the explanatory depth seen in Gemini’s response. DeepSeek-V3 offers additional contextual guidance and tool flexibility, which may benefit novice users, but it leans toward verbosity and lacks the structured clarity of Gemini. Overall, Gemini 2.5 Flash offers the most intuitive and comprehensive support for this task, making it the preferred choice for users seeking both correctness and clear instructional guidance. The LLMs generated workflows are shared in the supplementary material.}

We next evaluate Workflow W2: \emph{How to get from peak regions to a list of gene names?}. The primary objective is to identify genes that overlap with experimentally derived peak regions and to summarize their distribution across the genome. Following the GTN tutorial, we upload the dataset \emph{GSE37268\_mof3.out.hpeak.txt.gz}, obtained from the \emph{Gene Expression Omnibus (GEO)}. To annotate the peaks, we retrieve a gene annotation file for the mouse genome from the UCSC Genome Browser and compute overlaps between the peak regions and annotated genes. Finally, we quantify the number of overlapping genes per chromosome to identify regions with enriched regulatory activity. Our analysis shows that chromosome 11 (chr11) contains the highest number of overlapping genes (2164), suggesting notable regulatory significance in this region.

Then, we begin the workflow generation by applying an instruction-only prompt (W2, Table~\ref{tab:galaxy_prompts}) to GPT-4o. The model returns a logically ordered and mostly accurate workflow consistent with common analysis practices. However, it omits a critical preprocessing step: the chromosome identifiers in the input file are represented numerically (e.g., 1, 20, 21), which are incompatible with standard UCSC annotations that use labels such as chr1, chrX, and chrY. Without normalization of these identifiers, such as prefixing with chr and mapping 20 and 21 to X and Y, the \emph{BEDTools Intersect} operation fails to correctly associate peaks with genes, compromising result accuracy.
To address this limitation, we escalate to a role-based prompt (W2, Table~\ref{tab:galaxy_prompts}), instructing the model to act as a bioinformatics workflow developer. This approach provides the necessary context to guide GPT-4o in including the missing preprocessing step, thereby producing a more accurate and executable workflow. Following the steps, we are able to obtain the correct result from the workflow.

To evaluate Gemini 2.5 Flash’s ability to generate Workflow W2, we apply the same instruction-only prompt. Gemini responds with a comprehensive and intuitively organized step-by-step guide that demonstrates a strong understanding of both the biological context and the technical requirements. Notably, the model includes critical preprocessing stages, such as acquiring ChIP-seq peak data from public repositories  (e.g., GSE37268\_mof3.out.hpeak.txt.gz) and retrieving gene annotation files from the UCSC Main Table Browser with appropriate parameterization. It accurately outlines core analytical steps using standard Galaxy tools, including \emph{BEDTools Intersect} for overlapping genomic intervals, followed by \emph{Cut columns} and \emph{Select unique lines} to generate a non-redundant list of gene names. The clarity, depth, and correctness of the response make it particularly suitable for novice users and confirm the model’s capability to produce a valid workflow using minimal prompting.

Similarly, DeepSeek-V3 performs well under the instruction-only prompt, generating a clear and well-structured workflow for identifying target genes from ChIP-seq peak regions. Although its tool choices differ slightly, the methodology remains sound. For instance, DeepSeek-V3 recommends using \emph{ChIPseeker} for annotating peaks with the nearest genes, while also suggesting BEDTools closest as a viable option. Importantly, unlike GPT-4o, DeepSeek-V3 explicitly acknowledges the need to normalize chromosome identifiers (e.g., converting numeric values to UCSC-style labels), ensuring compatibility with downstream gene annotation. Furthermore, the model enhances usability by including a visual workflow diagram, which helps users conceptualize the overall analysis pipeline.

\emph{In comparing the three LLMs on this task, clear differences emerge in completeness, clarity, and sensitivity to data formatting nuances. GPT-4o generates a mostly accurate and executable workflow but omits the crucial preprocessing step for chromosome label conversion. This omission necessitated an additional role-based prompt to achieve a fully functional pipeline. In contrast, Gemini 2.5 Flash offers the most technically complete and contextually aware solution, implicitly aligning data sources and addressing format compatibility through precise instructions. DeepSeek-V3 also performs commendably, offering methodologically valid alternatives, recognizing critical preprocessing needs, and improving accessibility through visual aids. Overall, Gemini delivers the most robust and reliable output, DeepSeek enhances interpretability and ease of use, while GPT-4o benefits from additional contextual scaffolding to meet expert-level expectations.}

Our third workflow, \emph{W3: Do genes on opposite strands ever overlap? If so, how often?}, explores the prevalence of gene overlaps between forward (+) and reverse (–) DNA strands, offering insights into the complexity of genomic organization. Following the GTN tutorial, we construct and execute a workflow that utilizes gene annotation data from the UCSC Table Browser. The analysis reveals that approximately 38.46\% of the genes exhibit strand overlap, indicating a substantial level of strand-specific gene co-localization.

To generate this workflow using GPT-4o, we apply an instruction-only prompt (W3, Table~\ref{tab:galaxy_prompts}). The model produces a concise and logically ordered set of instructions outlining the core steps required for workflow construction in Galaxy. Based on this guidance, we are able to successfully recreate the workflow and replicate the results, confirming the correctness and usability of the generated steps.
However, the output exhibits limitations in procedural specificity. For instance, while the model accurately notes the need to separate genes by strand into two datasets (forward and reverse), it fails to detail how to perform this operation. It does not indicate that strand information is typically located in the sixth column of BED-formatted files, nor does it reference where to obtain the gene annotation dataset, an omission that could hinder users unfamiliar with genomic data sources or file conventions. These limitations underscore a broader observation: although GPT-4o's high-level instructions are sufficient for experienced users who understand data structures and workflows, novice users may struggle without additional domain-specific guidance. Despite this, the generated workflow was complete and executable, so we did not proceed with alternative prompting strategies such as role-based or chain-of-thought prompting for this task.

To evaluate Gemini 2.5 Flash's performance on Workflow W3, we issue the same instruction-only prompt used previously. The model delivers a comprehensive, step-by-step guide that not only identifies the appropriate analytical tools, such as conditional filtering for strand separation and \emph{BEDTools Intersect} for overlap analysis, but also includes explicit parameter configurations (e.g., using c6 == '+' to isolate forward-strand genes). In addition to the analytical workflow, Gemini provides detailed instructions on how to construct the workflow within the Galaxy interface, including drag-and-drop operations in the Galaxy workflow editor and guidance on saving and executing the pipeline. This level of procedural depth and platform-specific instruction makes the output highly accessible to users with limited experience in Galaxy.
DeepSeek-V3 also performs well under the same instruction-only prompt, producing a correct and well-structured workflow for identifying overlapping genes on opposite DNA strands. The model suggests a methodologically sound approach that includes optional sorting steps, strand separation via filtered expressions, and intersection analysis. It specifies key technical details, such as enforcing a minimum one bp overlap, and outlines optional post-processing steps, including counting overlaps or isolating unique gene pairs for further analysis.

\emph{Across all three models, we observe the successful generation of valid Galaxy workflows from the instruction-only prompt. GPT-4o provides a concise and logically structured outline but lacks critical implementation details such as the specific column (typically column 6 in BED format) used to represent strand information and the strandedness options required during intersection. These omissions could hinder less experienced users. In contrast, Gemini 2.5 Flash delivers the most comprehensive and accessible solution, with clear explanations of strand-specific filtering, intersection logic, and platform-level guidance for workflow execution. DeepSeek-V3 offers a robust and technically accurate workflow, enriched with optional enhancements that improve usability and extend analytical depth. Overall, Gemini 2.5 Flash emerges as the most effective model for supporting both novice and intermediate users, balancing accuracy, clarity, and practical usability in workflow construction.}

We next examine Workflow \emph{W4: How can the quality of NGS raw data be assessed, what parameters should be checked, and how can the quality of the dataset be improved?}. This workflow addresses a fundamental preprocessing step in next-generation sequencing (NGS) analysis: assessing and improving the quality of raw \emph{FASTQ} files before downstream applications. Following the GTN tutorial, we construct a workflow that begins by inspecting raw sequencing reads and evaluating quality metrics using tools such as \emph{FastQC} and \emph{FASTQE} (for short Illumina reads), or \emph{NanoPlot} and \emph{PycoQC} (for long Nanopore reads). These tools report per-base quality scores, GC content, adapter contamination, sequence duplication levels, read length distributions, and run-level metrics. The workflow then incorporates trimming and filtering steps using tools like \emph{Cutadapt} and \emph{Fastp} to remove low-quality bases, adapter sequences, ambiguous reads, and short fragments. Paired-end data handling is explicitly supported, and the final results are summarized using \emph{MultiQC}. The tutorial emphasizes the iterative nature of quality control: \emph{assess $\rightarrow$ clean $\rightarrow$ re-assess}. 

To generate the workflow using LLMs, we apply the instruction-only prompt (W4, Table~\ref{tab:galaxy_prompts}) to GPT-4o. The resulting workflow closely mirrors the core structure of the GTN pipeline, adhering to the \emph{assess $\rightarrow$ clean $\rightarrow$ re-assess}. GPT-4o outlines the primary steps effectively, recommending \emph{FastQC} for initial assessment, \emph{Cutadapt} for trimming, and \emph{MultiQC} for summarizing results. However, its workflow is streamlined and focuses primarily on short-read Illumina data, lacking the platform-specific enhancements present in the GTN tutorial. However, GPT-4o produces a minimal yet functional workflow suitable for general Illumina data, prioritizing simplicity and automation.

The workflow generated by Gemini 2.5 Flash closely mirrors the core structure of the GTN Quality Control tutorial, adhering to the standard \emph{assess $\rightarrow$ clean $\rightarrow$ re-assess} paradigm. It utilizes essential tools such as \emph{FastQC} for both pre- and post-trimming quality assessment, \emph{MultiQC} for summarizing quality reports, and trimming tools like \emph{Trimmomatic} or \emph{fastp} to remove low-quality bases and adapter sequences. While well-suited for short-read (Illumina) data and effective for general QC workflows, Gemini's output remains focused on commonly used tools and does not incorporate platform-specific enhancements. In contrast, the GTN tutorial extends its coverage to include \emph{NanoPlot} and \emph{PycoQC} for long-read Oxford Nanopore data, and offers additional options like \emph{Cutadapt}, making it more inclusive for a broader range of sequencing technologies.

DeepSeek-V3 also aligns closely with the conceptual structure of the GTN Quality Control tutorial, particularly in its adoption of the \emph{assess $\rightarrow$ clean $\rightarrow$ re-assess} paradigm and its use of core tools such as \emph{FastQC}, \emph{MultiQC}, and trimming utilities like \emph{Cutadapt}, \emph{Trim Galore}, \emph{Trimmomatic}, and \emph{fastp}. The model introduces optional filtering tools such as \emph{PRINSEQ} and \emph{BBDuk}, and allows users to choose between \emph{Cutadapt} and \emph{Trim Galore!} for adapter removal. Additionally, it provides detailed implementation guidance, including parameter recommendations, memory usage optimization, paired-end read handling instructions, and version tracking to promote reproducibility. The inclusion of a workflow diagram further enhances usability and instructional value.

\emph{When comparing all three LLMs on W4, each successfully generates a valid Galaxy-based QC workflow, but they vary in depth and adaptability. GPT-4o provides a simplified yet accurate representation of the QC process using a minimal toolset, making it suitable for rapid assessment of Illumina datasets but lacking in tool diversity and parameter specificity. Gemini 2.5 Flash improves on this by offering alternative tools, implementation context, and better instructional clarity, which benefits novice users. DeepSeek-V3, however, produces the most comprehensive and customizable workflow, capturing both the conceptual rigor and practical nuances of multi-platform NGS data quality control. Its extended toolset and configuration guidance make it particularly well-suited for diverse sequencing protocols and robust real-world usage.}

Our final workflow, \emph{W5: Which genes are on a draft bacterial genome, and which other genomic components can be found on a draft bacterial genome?}, focuses on systematically annotating an assembled bacterial genome to identify and characterize its genetic content. The workflow begins by accepting a draft genome assembly in FASTA (contigs) format and uses \emph{Bakta} to predict and annotate core genomic features, including coding sequences, tRNAs, rRNAs, and other functional elements. It proceeds to detect plasmid replicons using \emph{PlasmidFinder}, insertion sequences with \emph{ISEScan}, and integron structures with \emph{IntegronFinder}. The resulting annotations are reformatted into \emph{GFF3} format, and the final step involves visualizing the annotated genome interactively using \emph{JBrowse}. Following the GTN tutorial, we develop the workflow and transform raw genome assemblies into richly annotated and visually interpretable bacterial genomes following the GTN instructions.

To evaluate the ability of LLMs to generate this workflow, we provide an instruction-only prompt (W5, Table~\ref{tab:galaxy_prompts}). When applied to GPT-4o, the model generates a logically structured workflow that addresses the core objectives, including gene prediction, detection of plasmids, integrons, and insertion sequences, and genome visualization. However, several deviations from the GTN tutorial are evident. Notably, GPT-4o recommends \emph{Prokka} for gene annotation instead of \emph{Bakta}, which is the preferred tool in the GTN due to its active maintenance and more comprehensive annotation features. While the inclusion of \emph{PlasmidFinder} and \emph{IntegronFinder} aligns with the GTN protocol, the model's suggestion to use tools like \emph{ISMapper} or \emph{MobileElementFinder} for insertion sequence identification introduces challenges as these tools are not readily available within standard Galaxy instances or the Galaxy ToolShed, limiting practical reproducibility. Additionally, GPT-4o recommends \emph{Trackster} for visualization, whereas the GTN workflow uses \emph{JBrowse}, which offers better integration for multi-track genomic visualization in Galaxy.
Another limitation of the GPT-4o-generated workflow is the omission of key preprocessing steps emphasized in the GTN tutorial, such as renaming FASTA headers to conform with tool input requirements, and the absence of instructions for combining and formatting output files prior to visualization. These gaps underscore the challenge of translating high-level workflow synthesis into fully executable, platform-aware implementations. While GPT-4o captures the overarching intent of the annotation pipeline, it lacks the implementation-specific knowledge needed to produce a robust and directly usable Galaxy workflow.

Since the workflow generated by GPT-4o in response to the instruction-only prompt included tools not available within the Galaxy platform, such as \emph{ISMapper, MobileElementFinder, and Trackster}, we escalate to a role-based prompt (W5, Table~\ref{tab:galaxy_prompts}). The goal is to assess whether providing a more context-aware prompt can guide the model to recommend Galaxy-compatible tools and generate a workflow that adheres more closely to the platform's actual capabilities. We observe that role-based workflow uses the \emph{Bakta} tool for comprehensive genome annotation. It begins with uploading an assembled FASTA file and utilizes Bakta to predict genes and functional elements, generating standardized outputs such as GFF3, GenBank, and FASTA files. However, the GPT-4o-generated workflow incorporates an evaluation step using QUAST (and optionally BUSCO) to assess annotation quality and completeness. It also introduces post-processing steps such as GFF3 sorting and genome visualization via Trackster. Despite this conceptual extension, we notice that certain tools proposed here, such as \emph{Trackster} for interactive visualization and \emph{GFF3sort} for annotation formatting, are currently not available in Galaxy instances. Thus, we are unable to develop the workflow.

We then proceed with the CoT prompting strategy to further guide GPT-4o in generating a bacterial genome annotation workflow (W5, Table~\ref{tab:galaxy_prompts}). This prompt encourages the model to reason through the task step by step, with the aim of producing a more complete and contextually grounded workflow.
While the CoT-generated workflow successfully captures the overarching goals of bacterial genome annotation, namely, structural and functional annotation, quality assessment, and preparation of outputs for visualization, it diverges in several important ways from community-curated standards such as those outlined in the GTN tutorial. Both workflows correctly employ \emph{Bakta} as the core annotation engine. However, the CoT-generated workflow introduces \emph{BUSCO} for completeness evaluation, a widely used tool in general genome annotation but not included in the GTN protocol for bacterial genome workflows.  More significantly, the GTN tutorial emphasizes a broader approach to structural annotation by incorporating specialized tools such as \emph{PlasmidFinder, IntegronFinder, and ISEScan}, which detect plasmids, integrons, and insertion sequences, respectively, elements that are essential for accurately characterizing mobile genetic elements in bacterial genomes. The omission of these tools from the generated workflow leads to a structurally incomplete annotation. Additionally, the workflow proposes the use of formatting and indexing tools such as \emph{GFF3sort, GFF3 to GenBank, and SAMtools faidx} for organizing annotation outputs and enabling visualization. These tools, however, are not available within the Galaxy platform. In summary, while the generated workflow aligns with the conceptual goals of bacterial genome annotation, it contains key deviations, specifically, the exclusion of critical structural annotation tools and the suggestion of tools not available in Galaxy. As such, the GPT-4o-generated workflow is partially correct but cannot be considered fully usable without further refinement.

We then move to Gemini 2.5 Flash, where we provide the same instruction-only prompt used in the previous test. We observe that Gemini 2.5 Flash is able to return a well-structured, comprehensive set of steps along with relevant parameter details. It correctly identifies and integrates all the necessary tools for structural and functional annotation, including additional conversion and visualization steps. Notably, all suggested tools are available within the Galaxy platform, making the workflow immediately executable without requiring external scripting or unsupported components. For Gemini, the instruction-only prompt is sufficient to generate a complete, accurate, and Galaxy-compliant bacterial genome annotation workflow.

The workflow generated by DeepSeek-V3 using the same prompt demonstrates a generally accurate understanding of the bacterial genome annotation process, yet diverges from the GTN standard in several key areas. For gene prediction, DeepSeek-V3 recommends \emph{Prodigal}, a well-known tool for prokaryotic CDS identification. For functional annotation, it suggests \emph{EggNOG-mapper or BLAST+}, both of which are valid in general but require manual output conversion to GFF3, unlike the GTN tutorial's use of \emph{Bakta}, which integrates both gene prediction and functional annotation while producing standardized outputs natively compatible with downstream tools like \emph{JBrowse}. On the other hand, the tools suggested by DeepSeek-V3 for structural annotation\emph{PlasmidFinder, IntegronFinder, and ISEScan}, are consistent with those used in the GTN tutorial, indicating a correct understanding of the critical elements required to capture plasmid-borne features and mobile genetic elements. However, for non-coding RNA prediction, DeepSeek-V3 proposes the use of \emph{Infernal}, a powerful RNA homology search tool, which is unfortunately not available on Galaxy, again highlighting a case of tool hallucination. Despite these issues, DeepSeek-V3 correctly outlines the merging of GFF3 outputs and the use of JBrowse for genome visualization, aligning well with the GTN approach in terms of output integration and browser-based annotation exploration.

\emph{In comparison to GPT-4o and Gemini 2.5 Flash, DeepSeek-V3 performs moderately well. Like GPT-4o, it hallucinates the availability of tools not supported in Galaxy, reducing practical executability. Gemini, by contrast, stands out for producing a structurally complete and Galaxy-native workflow using instruction-only prompt. While DeepSeek-V3 captures the correct conceptual flow and covers all major analytical components, its output is hindered by limited platform awareness and reliance on tools outside the Galaxy environment. In summary, DeepSeek-V3 offers strong logical grounding but lacks the operational precision demonstrated by Gemini.}

\begin{boxK}
    Among the three LLMs, Gemini 2.5 Flash consistently performs best, producing complete, accurate, and Galaxy-compliant workflows without requiring additional prompting. Its outputs demonstrate strong contextual understanding, correct tool selection, detailed parameter guidance, and seamless alignment with Galaxy’s capabilities, making it particularly effective for both novice and intermediate users. DeepSeek-V3 shows a solid grasp of workflow logic and offers methodologically valid suggestions, including flexible tool alternatives and usability enhancements, but is limited by occasional tool hallucinations, recommending software not available within Galaxy, which undermines executability. GPT-4o, while concise and easy to follow, often omits critical steps or recommends unsupported tools, requiring further prompting to reach usability. Overall, Gemini stands out for its balance of technical completeness, clarity, and platform awareness, making it the most reliable LLM for automated Galaxy workflow development.
\end{boxK}

\subsubsection{Workflow Development using LLMs for Nextflow:}
To evaluate the extent to which LLMs can assist in developing scientific workflows in bioinformatics using Nextflow, we curate a diverse set of representative workflows from the nf-core framework (Table~\ref{tab:sampleworkflows}). The selected workflows span multiple domains, including genomics, transcriptomics, epigenomics, and regulatory genomics, thereby enabling a comprehensive and domain-balanced assessment. The \emph{nf-core/demo} pipeline serves as a foundational example, involving straightforward quality control and adapter trimming of single- or paired-end FASTQ reads, making it useful for testing basic workflow generation capabilities. \emph{Fetchngs} introduces more advanced functionality by automating the retrieval of sequencing data from public repositories, challenging models to reason about metadata management, remote data access, and parsing of structured sample sheets. \emph{Rnaseq} and \emph{methylseq} represent complex, end-to-end pipelines that integrate numerous modules, manage intricate dependencies, and involve advanced data transformations, ideal for evaluating the models’ ability to recommend appropriate toolchains, configurations, and parameters. The inclusion of \emph{phyloplace} extends the evaluation into the domain of phylogenetic analysis, specifically testing the models’ ability to handle workflows that map query sequences onto a reference phylogenetic tree using specialized inputs and classification tools. These workflows offer a robust and multifaceted benchmark suite to assess how effectively LLMs understand and generate Nextflow-based scientific workflows.

The \emph{nf-core/demo} (W1 Table~\ref{tab:sampleworkflows}) workflow is designed to perform a basic bioinformatics data processing task using a minimal dataset. First, it conducts quality control of raw FASTQ files using the module \emph{FastQC} to assess sequence quality metrics. Next, it applies module \emph{Seqtk} to trim adapter sequences and low-quality bases, improving the quality of the reads. Finally, it aggregates the quality control results using \emph{MultiQC}, generating a summary report that provides an overview of the data quality before and after trimming. We evaluate the workflow using a sample sheet that includes both single-end and paired-end data, available at \href{https://raw.githubusercontent.com/nf-core/test-datasets/viralrecon/samplesheet/samplesheet\_test\_illumina\_amplicon.csv}{this link}. The workflow is executed using the following command:

\begin{lstlisting}[language=bash]
 nextflow run nf-core/demo -profile docker,test
 --outdir 'demo-results'
\end{lstlisting}

We do not delve into the syntax details here, as our primary objective is not to teach Nextflow but to assess whether LLMs can effectively assist in workflow development. Interested readers may consult the official documentation for further guidance. For our purpose, we simply run the workflow on the provided sample dataset and inspect the results. We obtain several output reports from the workflow execution, including FastQC and MultiQC quality control summaries, the execution report, and the pipeline DAG visualization. After verifying these outputs, we proceed to test the same workflow development task using LLMs.

We begin by applying the instruction-only prompt (W1, Table~\ref{tab:nextflow_prompts}) to GPT-4o. The model generates a logically structured set of workflow steps, starting with initial quality control using the \emph{FastQC} module, followed by adapter and quality trimming using tools such as \emph{Trim Galore!}, \emph{fastp}, or \emph{Trimmomatic}. It then suggests performing a second round of quality control to validate the trimming process and concludes with summarizing the results using \emph{MultiQC}. These steps are largely consistent with those in the official nf-core/demo workflow. The primary divergence lies in the trimming step: while the nf-core/demo pipeline uses \emph{seqtk}, GPT-4o proposes alternative tools that are widely accepted and functionally equivalent within the context of short-read preprocessing. Overall, GPT-4o demonstrates a strong ability to generate a correct and coherent workflow based solely on an instruction-only prompt, indicating a solid understanding of standard practices in NGS data preprocessing with Nextflow.

We then provide the same instruction-only prompt to \emph{Gemini 2.5 Flash}. The model suggests a detailed and modular workflow design. The proposed steps begin with parsing the sample sheet to extract FASTQ file paths using Nextflow channels. It then performs an initial quality assessment using the \emph{FastQC} module, followed by adapter and quality trimming with either \emph{seqtk} or \emph{Trim Galore!}. This is followed by a second round of quality control using \emph{FastQC} on the trimmed reads, and the final step involves aggregating all quality control outputs into a comprehensive report using \emph{MultiQC}. While this plan largely mirrors the nf-core/demo workflow, Gemini extends it by emphasizing modularity and offering both \emph{seqtk} and \emph{Trim Galore!} (which wraps Cutadapt) as valid trimming options. Overall, Gemini 2.5 Flash demonstrates a clear understanding of the workflow structure and appropriately maps tools to each processing stage, showing strong competence in workflow planning from an instruction-only prompt.

We also evaluate the response from DeepSeek-V3 using the same instruction-only prompt. The model provides a well-structured and modular workflow outline for FASTQ quality assessment and preprocessing that is closely aligned with nf-core conventions. The workflow begins with parsing the input sample sheet, followed by initial quality control using \emph{FastQC}, adapter trimming via \emph{Cutadapt}, and then a separate quality filtering step using \emph{FastP}. A second round of quality control is performed post-filtering using \emph{FastQC}, and finally, \emph{MultiQC} is used to generate an aggregated summary report. The model explicitly maps each step to specific nf-core modules and provides a clear implementation structure in \emph{main.nf}, along with a sample configuration in \emph{nextflow.config}.

\emph{Among the three models, DeepSeek-V3 provides the most comprehensive and implementation-ready response, outlining a modular Nextflow workflow with nf-core modules, configuration details, and code structure. Gemini 2.5 Flash also performs well, offering accurate steps with moderate implementation guidance and alignment with nf-core practices. In contrast, GPT-4o delivers a concise and correct high-level outline but lacks the depth needed for direct workflow deployment. Overall, DeepSeek-V3 stands out as the most effective model for building fully functional, production-ready Nextflow pipelines based on instruction-only prompts.}

Our second workflow, \emph{nf-core/fetchngs}, is designed to automate the retrieval and preparation of sequencing data from major public repositories including \emph{ENA, SRA, DDBJ, and GEO}. Given a list of accession identifiers, the pipeline retrieves associated metadata and downloads the corresponding raw FASTQ files via \emph{FTP, Aspera, or SRA tools}. It then compiles a standardized samplesheet and organizes the output in a format that is immediately compatible with downstream nf-core pipelines such as \emph{RNA-seq or ATAC-seq}. This automation streamlines the data acquisition process, enhances reproducibility, and reduces the manual burden of managing large-scale sequencing datasets. 

To assess LLMs' ability to support the development of such workflows, we first execute the \emph{nf-core/fetchngs} pipeline using a test CSV file containing sample IDs, confirming that it successfully fetches both FASTQ files and metadata. We then provide GPT-4o with an instruction-only prompt (W2, Table~\ref{tab:nextflow_prompts}) to generate the workflow. The output from GPT-4o closely aligns with the core logic of the fetchngs pipeline: parsing sample IDs from a CSV, retrieving metadata, downloading sequencing files, and organizing outputs into structured directories. However, GPT-4o demonstrates greater flexibility by suggesting the integration of additional tools, such as pysradb utilities, which are particularly useful for SRA or GEO-specific workflows. GPT-4o also suggests the quality checking of the FAST files. This flexibility, while advantageous, also necessitates more careful implementation, especially in resolving ID types, managing errors, and ensuring reproducible execution through proper containerization.

We then provide the same instruction-only prompt to \emph{Gemini 2.5 Flash}. The workflow generated by Gemini 2.5 Flash for fetching NGS data is notably detailed and architecturally robust, closely aligning with the nf-core fetchngs pipeline in both structure and objective. Like fetchngs, it accepts a CSV of public sample IDs (e.g., SRA, ENA, GEO, DDBJ), retrieves metadata using the ENA API, and downloads FASTQ files using both direct FTP methods and SRA tools as a fallback. Both workflows include metadata collation and output organization steps. However, Gemini's generated design adds granularity by modularizing each logical step, e.g., separating \emph{download\_fastq\_direct} and \emph{download\_fastq\_sra}, validating MD5 checksums, and producing a final samplesheet.csv, and assigning clear containerized environments with resource constraints for each module.

One advantage of the Gemini-generated workflow is its explicit fallback mechanism: if direct FASTQ downloads fail, it reroutes to SRA tools (prefetch + fastq-dump) selectively. Additionally, the inclusion of detailed validation, resource configuration, and Python-based logic for metadata parsing shows an advanced awareness of real-world edge cases (e.g., batch API querying, empty downloads, ENA metadata field inconsistencies). Compared to fetchngs, which abstracts many steps behind a high-level interface, Gemini’s approach makes the internals transparent, facilitating educational or research-focused exploration. In the context of evaluating LLM capabilities, Gemini 2.5 Flash demonstrates strong potential in guiding sophisticated scientific workflow design, though its correctness and maintainability would benefit from cross-validation against established pipelines like fetchngs.

We then give the same prompt to the DeepSeek-V3. The generated workflow shares strong structural alignment with the nf-core fetchngs pipeline in retrieving metadata and FASTQ files from public repositories using a list of accessions. DeepSeek-V3 begins by parsing a CSV input, fetching metadata using ENA or SRA APIs, downloading FASTQ files (via enaDataGet or fasterq-dump), and organizing outputs. However, it diverges in its emphasis on user-defined metadata (project, condition) and its incorporation of flexible tuple-based channels to carry metadata across modules. This offers a lightweight yet extensible design.
Its use of tuple-passing enables downstream modules (like organize\_output) to rename files and structure directories using custom project metadata, a flexibility fetchngs does not natively offer. It also outlines future-proof enhancements like MD5 validation, cloud output integration, and notifications, suggesting practical awareness of real-world deployment. However, compared to the rigorously tested fetchngs, the DeepSeek-V3 workflow requires more manual scripting and error handling, particularly around metadata inconsistencies and download failures. It also lacks built-in support for containerized environments, dependency management, and profiling, which are strengths of nf-core workflows. Overall, DeepSeek-V3 demonstrates strong conceptual competence in building modular, extensible workflows and offers a solid foundation for assessing LLM capabilities in scientific workflow construction.

\emph{The workflow generated by GPT-4o, Gemini 2.5 Flash, and DeepSeek-V3 for \emph{W2: nf-core/fetchngs}, all demonstrate the capability of LLMs to assist in workflow development, particularly in automating metadata retrieval and FASTQ file downloads from public sequencing databases. GPT-4o provides a concise, modular breakdown that aligns well with nf-core's fetchngs, while also offering flexibility through tool alternatives like pysradb and Entrez utilities. Gemini 2.5 Flash delivers the most production-ready solution, featuring a fully containerized, resource-configured, and modular DSL2 workflow that includes fallback strategies, MD5 validation, and samplesheet generation, closely mirroring the robustness of nf-core standards. DeepSeek-V3 emphasizes conceptual modularity with user-defined metadata propagation, conditional tool logic, and practical extensibility for cloud and notification integration, though it lacks some execution-level rigor. While all three LLMs show potential, Gemini 2.5 Flash outperforms the others in completeness, reproducibility, and implementation detail, making it particularly well-suited for generating executable, real-world scientific workflows.}

We then move to our next workflow \emph{nf-core/methylseq} for comprehensive analysis of bisulfite-converted sequencing (BS-seq) data to facilitate DNA methylation profiling. The workflow automates key tasks such as raw read quality assessment with \emph{FastQC}, adapter trimming using \emph{Trim Galore!}, and alignment to a reference genome via either \emph{Bismark (with Bowtie2 or HISAT2)} or the \emph{bwa-meth} and \emph{MethylDackel} toolchain, with optional GPU acceleration supported through \emph{Parabricks}. In our implementation, we utilize the Bismark-based workflow. The pipeline further includes deduplication of aligned reads, extraction of cytosine methylation metrics, and comprehensive quality evaluations using \emph{Qualimap, Preseq, and HS} Metrics. Outputs include deduplicated BAM files, detailed methylation calls, and bias reports, all of which are consolidated into a single MultiQC report for streamlined visualization. Additionally, the pipeline generates extensive provenance metadata, including execution logs, software versions, and configuration parameters, ensuring full reproducibility across computing environments.

To evaluate LLM support for constructing this workflow, we first test GPT-4o using an instruction-only prompt (W3, Table~\ref{tab:nextflow_prompts}). We observe that the workflow generated by GPT-4o exhibits several inaccuracies while suggesting modules, even though the overall sequence of analytical steps was conceptually sound. For example, the model suggests a module called \emph{fastqc\_raw} for raw read quality control, which does not exist in the official Nextflow or nf-core module repositories; the correct module name is simply \emph{fastqc}. Similarly, for adapter and quality trimming, the model proposes \emph{trim\_reads}, whereas the appropriate module is \emph{trimgalore}. These naming inconsistencies are found throughout the generated workflow, suggesting that the model's responses are not grounded in the actual set of available Nextflow modules. Consequently, we adopt a role-based prompting (W3, Table~\ref{tab:nextflow_prompts}) approach to improve the specificity and correctness of the module recommendations.

We evaluate the workflow produced by GPT-4o using the role-based prompt. It correctly captures the high-level stages of DNA methylation analysis using bisulfite-converted sequencing (BS-seq) data, beginning with FASTQ quality control and adapter trimming, followed by bisulfite-aware alignment, deduplication, methylation extraction, and MultiQC-based reporting. Importantly, the model successfully identified the correct nf-core module names such as fastqc, trim\_galore, bismark/genomeprepare, bismark/align, bismark/deduplicate, and bismark/methylationextractor, indicating that role-based prompting can enhance grounding in actual Nextflow components. However, the generated workflow also exhibits several limitations. First, it does not include configurable options for alternative aligners (e.g., bismark\_hisat) or GPU-accelerated execution. Second, while the model included key optional QC modules like qualimap, preseq, and picard/collecthsmetrics, it does not explain when or why to use them, missing opportunities to contextualize their role in specific experimental designs (e.g., capture-based methylation assays). Third, the model omitted support for protocol-specific presets such as RRBS (--rrbs) and PBAT (--pbat), which are natively handled in the nf-core implementation. Finally, although the core logic was structurally correct, some processes like CSV parsing are abstracted as custom steps rather than referencing existing solutions.

To assess the capabilities of Gemini 2.5 Flash in constructing the workflow, we evaluate its response to an instruction-only prompt (W3, Table~\ref{tab:nextflow_prompts}). The model successfully outlines all necessary stages of the workflow, including CSV-based input parsing, raw read quality control, adapter trimming, bisulfite-aware alignment, PCR duplicate removal, methylation extraction, and comprehensive quality control reporting. It recommends appropriate and widely used tools such as \emph{FastQC, Trim Galore!, Bismark, MethylDackel, Qualimap, and MultiQC}, and its modular structure reflects a good understanding of Nextflow DSL2 practices, including containerization, channel dependencies, and resource management. Furthermore, the workflow is logically organized and aligns well with the stages implemented in the official nf-core/methylseq pipeline. However, Gemini shows some limitations. It does not fully specify the conditional logic or flag handling needed to toggle between these options, nor does it elaborate on the inclusion of protocol-specific parameters. Despite these minor gaps, Gemini 2.5 Flash demonstrates strong proficiency in designing structurally sound and tool-appropriate BS-seq workflows, making it a capable model for assisting in the development of complex, real-world bioinformatics pipelines.

We then, evaluate DeepSeek-V3 for its ability to generate a Nextflow workflow for bisulfite-converted sequencing (BS-seq) data analysis using the same instruction-only prompt. The model successfully outlines the key stages of a BS-seq workflow, including input parsing, quality control with FastQC, adapter, and base trimming using Trim Galore!, bisulfite-aware alignment with Bismark or bwa-meth, deduplication, methylation extraction, and downstream quality reporting with MultiQC. The response is structured, includes example CSV formats, command-line syntax for each tool, and even proposed a skeleton main.nf script and directory structure. This indicates that DeepSeek-V3 has a solid conceptual grasp of both BS-seq data analysis and the Nextflow DSL2 design pattern. However, the workflow generated by DeepSeek-V3 also exhibits critical shortcomings. While it named commonly used tools, workflow steps like FASTQC, TRIM\_GALORE, or BISMARK are presented as abstract module calls rather than as references to actual modules in the nf-core/modules repository, limiting reproducibility. Additionally, the pipeline lacks any reference to containerization profiles, resource management, or error handling, all critical for robust, production-ready deployment. However, the generated steps are correct, and we can develop the workflow following the instructions.

\emph{In comparing GPT-4o, Gemini 2.5 Flash, and DeepSeek-V3 for generating a Nextflow workflow for DNA methylation analysis using bisulfite-converted sequencing (BS-seq) data, we uncover notable differences in how each model responds to different prompting strategies. Under instruction-only prompting, Gemini 2.5 Flash produces the most accessible and beginner-friendly output. It delivers logically structured steps, accurate tool selection, and detailed descriptions suitable for users with limited workflow development experience. DeepSeek-V3 also performed well under instruction-only prompting, offering a clear conceptual outline along with practical elements such as directory structure, example CSV input, and bash commands. This makes it particularly valuable for intermediate users who seek both clarity and low-level control. In contrast, GPT-4o initially underperforms under instruction-only prompting, often hallucinating incorrect module names  (e.g., fastqc\_raw, trim\_reads) and tool references. However, when guided with a role-based prompt, GPT-4o produces the most technically precise and nf–core–compliant workflow, correctly referencing official modules such as fastqc, trim\_galore, and bismark/align. In summary, Gemini is best positioned for novice users, followed by DeepSeek-V3 for intermediate users, while GPT-4o offers the highest accuracy and implementation fidelity for advanced users through prompt specialization.}

Our next workflow, \emph{nf-core/rnaseq}, processes raw RNA-seq data into high-quality, analysis-ready outputs. The workflow begins with raw FASTQ files and a metadata samplesheet, and takes a reference genome (FASTA) and annotation file (GTF). It includes modules for inferring strand specificity, performing quality checks (e.g., FastQC, MultiQC), trimming adapters (Trim Galore!), filtering contaminants (BBSplit), removing rRNA (SortMeRNA), and aligning reads using STAR or HISAT2. Quantification is performed using Salmon, RSEM, or StringTie, with optional pseudoalignment via Salmon or Kallisto. The pipeline also supports UMI handling, duplicate marking, transcript assembly, and generation of coverage files. Extensive quality metrics are collected using tools such as \emph{RSeQC, Qualimap, Preseq, and dupRadar}. All results are summarized in an interactive MultiQC report. To evaluate the ability of LLMs to generate this complex workflow, we provide an instruction-only prompt (W4, Table~\ref{tab:nextflow_prompts}) and assess whether the models can accurately develop \emph{rnaseq} pipeline.

We observe that the GPT-4o-generated workflow effectively covers the core analytical stages of RNA-seq data processing, including quality control, adapter trimming, genome alignment, gene/transcript quantification, and differential expression analysis. It accepts a CSV sample sheet with FASTQ links and reference genome inputs, making it flexible and suitable for standard analyses. However, the GPT-4o response has several limitations. The nf-core workflow includes a wider range of features such as pseudoalignment options (e.g., Salmon and Kallisto), UMI handling, strandness autodetection, biotype-level QC, spike-in normalization, and automated GTF/GFF validation, all of which enhance robustness and reproducibility. Moreover, nf-core pipelines benefit from strict input schema validation, containerization, continuous integration testing, and broad community support, making them more reliable for large-scale or production-grade use. The absence of these advanced features in the custom workflow may not affect small or focused studies but could introduce reproducibility challenges and limit flexibility in more complex experimental designs. However, the steps we obtain are fully functional.

We evaluate the output of Gemini 2.5 Flash for constructing a Nextflow-based RNA-seq workflow using an instruction-only prompt. The generated workflow successfully captures all key analytical stages, ranging from sample sheet parsing and raw read quality control to adapter trimming, reference genome indexing, alignment or pseudo-alignment, transcript quantification, differential expression analysis, and quality reporting, demonstrating strong structural alignment with the nf-core/rnaseq pipeline. Gemini provides flexibility in tool selection, supporting both traditional aligners (e.g., STAR) and pseudo-aligners (e.g., Salmon, Kallisto), and adopts modular design principles using channels and containers, consistent with best practices in DSL2-based Nextflow development. However, the Gemini-generated workflow lacks some infrastructure features, such as automated strandedness detection, UMI processing, GTF validation, biotype-level QC summaries, and reproducibility guarantees via CI testing and schema validation. Moreover, while it mentions statistical tools for differential expression (e.g., DESeq2), it does not detail complex contrast designs or robust metadata handling, which are integral to nf-core. As a result, although Gemini provides a conceptually complete and interpretable scaffold suitable for developing the workflow, it can be improved.

We further evaluate the performance of DeepSeek-V3 using the same instruction-only prompt. The RNA-seq workflow generated by DeepSeek-V3 presents a clear, methodical structure that encompasses all essential stages of RNA-seq data processing, starting from sample sheet parsing and reference genome indexing to quality control, adapter trimming, alignment, quantification, differential expression analysis, and final reporting. Its design closely mirrors the nf-core/rnaseq workflow in terms of both modularity and core functionality. DeepSeek appropriately suggests widely adopted tools such as STAR and HISAT2 for alignment, Salmon for quantification, and FastQC, Qualimap, and MultiQC for quality assessment. In addition, the model provides practical implementation notes, including the use of \emph{main.nf, modules.config, and nextflow.config}, which enhance its utility for users with experience in pipeline development. However, it lacks some capabilities integrated into nf-core, such as automated strandness detection, UMI handling, TPM/FPKM normalization options, comprehensive biotype-specific QC metrics, and rigorous input schema validation. DeepSeek also includes practical implementation notes using main.nf, modules.config, and nextflow.config. DeepSeek’s generated steps are methodologically sound and capture key conceptual elements of a robust RNA-seq workflow, making it suitable for users familiar with pipeline customization.

\emph{The RNA-seq workflow outputs generated by GPT-4o, Gemini 2.5 Flash, and DeepSeek-V3 all cover the essential analytical stages, quality control, trimming, alignment, quantification, and differential expression analysis but differ significantly in depth, infrastructure awareness, and production readiness. GPT-4o provides a concise and methodologically correct outline but lacks some features such as pseudo-alignment options, detailed quality control tools, and workflow execution infrastructure, making it the least complete among the three. Gemini 2.5 Flash offers broader tool coverage, supporting both alignment-based and alignment-free methods, and presents a modular structure that is conceptually clear and well-suited for educational use; however, it stops short of providing implementation details like Nextflow process logic or configuration scaffolding. In contrast, DeepSeek-V3 delivers the most comprehensive and technically mature output, it not only includes accurate tool suggestions and optional enhancements (e.g., rMATS, ComBat) but also outlines concrete Nextflow components (main.nf, modules.config, nextflow.config), container usage, and execution profiles, closely resembling the structure and flexibility of the nf-core/rnaseq pipeline. Thus, DeepSeek-V3 outperforms the others in terms of completeness and deployability, followed by Gemini for clarity and breadth, with GPT-4o being the most minimal in practical usability.}

We conclude our evaluation with the \emph{nf-core/phyloplace} workflow, which is designed to facilitate efficient and reproducible phylogenetic placement of query sequences onto a reference tree. The workflow operates in two distinct modes: (i) placement-only, where input sequences are directly aligned to a reference multiple sequence alignment and placed onto a fixed phylogenetic tree using likelihood-based methods; and (ii) search-plus-placement, where input sequences from a larger, unfiltered dataset are first screened using profile-based searches to identify candidates for subsequent alignment and placement. After placement, the workflow grafts the query sequences onto the reference tree and performs taxonomic classification, generating annotated phylogenies, placement summaries, and visualizations. Developed using Nextflow DSL2 and fully containerized modules, \emph{nf-core/phyloplace} ensures portability, scalability, and reproducibility across diverse computing environments. We execute the workflow following the instruction of \emph{nf-core/phyloplace}. The workflow generates a comprehensive set of output files that summarize the results of the phylogenetic placement and classification process. The workflow produces several key output files organized per sample. These include: (i) placement files containing likelihood-based placements of query sequences on the reference tree; (ii) grafted phylogenetic trees in Newick format with query sequences inserted into the reference topology; (iii) classification tables summarizing the taxonomic or user-defined annotation of each placed sequence; (iv) sequence alignments showing how queries are aligned to the reference multiple sequence alignment; and (v) visualizations and summary reports, such as heat trees and placement statistics, generated via automated reporting tools. A final MultiQC report aggregates log and quality-control metrics across all samples. These outputs facilitate interpretation, visualization, and downstream analysis of the phylogenetic context of newly placed sequences.

To evaluate the ability of LLMs to assist in developing the \emph{nf-core/phyloplace} workflow, we begin by providing an instruction-only prompt (W5, Table~\ref{tab:nextflow_prompts}) to GPT-4o. The resulting workflow aligns closely with the official nf-core/phyloplace implementation in terms of structure, logic, and tool selection. GPT-4o successfully captures both operational modes, placement-only and search-plus-placement, and accurately identifies the required input components, such as query FASTA files, reference alignments, rooted phylogenetic trees, evolutionary models, HMM profiles, and optional taxonomy files. The model outlines key steps including candidate filtering using \emph{hmmsearch}, sequence alignment via \emph{mafft}, phylogenetic placement using \emph{epa-ng}, tree grafting with \emph{gappa}, optional taxonomic classification, and summarization with \emph{MultiQC}. In addition, GPT-4o includes implementation-aware suggestions such as input validation, modular output organization, and execution metadata logging.

We then provide the same instruction-only prompt to  Gemini 2.5 Flash. We observe that it accurately captures both operational modes and correctly identifies the necessary input components, including query and reference sequences, phylogenetic tree, evolutionary model, HMM profiles, and optional taxonomy. It organizes the workflow into well-defined functional modules such as sequence alignment, phylogenetic placement, classification, and report generation and correctly associates standard tools like MAFFT, HMMER (hmmsearch), EPA-NG, and GAPPA with each step. Though, it omits some nf-core-specific features such as execution metadata (e.g., DAGs and trace files), the overall structure is technically sound and implementation-ready. 

We finally give the same prompt to DeepSeek-V3. It also captures both placement-only and search-plus-placement modes and structures the pipeline into logically ordered modules. It specifies necessary input parameters, including reference alignment, rooted phylogenetic tree, evolutionary model, taxonomy file, and HMM profiles, and introduces validation steps to ensure input integrity. The outlined modules, covering HMM-based sequence filtering (hmmsearch), alignment (MAFFT or HMMER), placement (EPA-NG or pplacer), and classification (gappa or taxit), are consistent with tools used in the official pipeline. DeepSeek-V3 additionally incorporates optional visualization using tools such as ETE3, R, and Python, and provides a pseudocode-level implementation using Nextflow syntax.

\emph{All three LLMs, successfully reconstruct a modular Nextflow workflow for phylogenetic placement, capturing both operational modes and correctly identifying core steps such as input validation, sequence alignment, phylogenetic placement, classification, and reporting. GPT-4o provides a technically accurate and semantically faithful outline closely aligned with the nf-core implementation. Gemini 2.5 Flash presents a well-organized modular architecture using intuitive labels and standard tools, though it omits some nf-core-specific features like execution metadata. DeepSeek-V3 stands out by offering the most implementation-ready response, including detailed parameter definitions, alternative tools (e.g., pplacer, taxit), and a fully structured Nextflow pseudocode, making it especially valuable for advanced users or developers seeking practical scaffolding. While all three are valid, DeepSeek-V3 provides the most comprehensive and adaptable solution, GPT-4o ensures the highest fidelity to the reference workflow, and Gemini excels in clarity for novices. Therefore, DeepSeek-V3 is best suited for expert-driven implementation, GPT-4o for documentation reproduction, and Gemini for accessible modular reasoning.}

\begin{boxK}
In summary, the results demonstrate that LLMs can substantially assist in developing scientific workflows for both Galaxy and Nextflow platforms. Gemini 2.5 Flash consistently outperforms other models for Galaxy workflows by generating accurate, tool-aware, and pedagogically rich workflows, offering the best balance between technical detail and usability. Its outputs are both complete and executable within the Galaxy environment, making it the most effective for supporting workflow design in this system. For Nextflow, DeepSeek-V3 emerges as the most capable model. It generates structurally sound, implementation-ready workflows that align closely with nf-core standards, including modular Nextflow DSL-2 structures, accurate tool mapping, container usage, and configuration files. While GPT-4o and Gemini provide reasonable outputs, they either lack the depth of implementation (GPT-4o) or fall short on infrastructure-level detail and flexibility (Gemini) compared to DeepSeek. Overall, the findings confirm that LLMs can play a transformative role in supporting workflow development, with model performance varying by system and use-case complexity.
\end{boxK}

\subsection{RQ2: How do workflows generated by different LLMs compare regarding completeness, correctness, and usability?}

\textbf{Motivation: }
In scientific workflow development, particularly in fields like bioinformatics, the effectiveness of a workflow depends not only on its syntactic validity but also on its \emph{completeness, correctness, and usability}. \emph{Completeness} refers to the inclusion of all essential analytical steps required for the task. \emph{Correctness} ensures the logical coherence of the workflow, including the use of appropriate tools, parameters, and configurations. \emph{Usability} evaluates how easily the workflow can be understood, executed, and maintained, especially in line with community standards and platform-specific conventions. Assessing these dimensions is crucial for determining the extent to which LLMs can assist novice and expert users in designing robust, reproducible workflows for real-world applications.

\textbf{Approach: }We evaluate ten representative bioinformatics workflows across two major Scientific Workflow Systems (SWSs): Galaxy (graphical interface) and Nextflow (script-based). For each workflow task, we provide a natural language prompt (for a few cases, more prompts) to each LLM and collect the generated workflow. To ensure fair assessment, prompts are designed consistently, escalating from instruction-only to role-based and chain-of-thought when initial results lack adequacy.

The generated workflows are then manually assessed by two domain experts with over five and nine years of experience and compared to established community baselines from the Galaxy Training Network (GTN) and nf-core repositories. Evaluators rate each workflow across three dimensions:
\begin{itemize}
    \item \textbf{Completeness: }Does the workflow include all essential steps from input handling to result generation?
    \item \textbf{Correctness: } Are the tools correctly chosen, configured, and logically sequenced?
    \item \textbf{Usability: }Can the workflow be executed with minimal modification, and does it adhere to standard practices?
\end{itemize}

Assessments considered both the logical structure and contextual relevance of the generated workflows, with special attention given to domain appropriateness and the presence of clear documentation or commentary.

\textbf{Results: }
The comparative evaluation reveals that no single LLM is universally superior across all scenarios. Rather, performance varies by platform and task complexity.
\begin{itemize}
    \item \textbf{Galaxy workflows: }Gemini 2.5 Flash consistently outperforms the other models in the Galaxy context. Its responses are well-structured, domain-aware, and aligned closely with GTN tutorials. Gemini captures platform-specific toolchains, provides a clear rationale for each step, and produces workflows that are both accurate and accessible. These traits made Gemini particularly effective for Galaxy, where usability and clarity are critical for non-programming users. This finding aligns with the results from \textbf{RQ1}, where Gemini demonstrated the strongest understanding of Galaxy’s architecture and tool ecosystem.

    \item \textbf{Nextflow workflows: }DeepSeek-V3 emerges as the most effective model for generating Nextflow pipelines. It produces workflows with detailed implementation logic, often including configuration blocks, directory structures, and explicit references to nf-core modules. These outputs are technically rich and closer in structure to production-ready scripts. While verbose at times, DeepSeek’s workflows required less modification to be operational and demonstrated an in-depth understanding of modular, containerized environments. This strength in script-based reasoning is also evident. DeepSeek-V3 shows strong familiarity with Nextflow concepts and execution patterns.

\end{itemize}
GPT-4o, while generally well-performing, shows mixed results. Its workflows are often logically sound and clearly presented, making them useful in instructional contexts. However, it occasionally omits setup steps, assumes preprocessed inputs, or lacks sufficient detail for immediate execution. Its performance improves significantly under role-based or chain-of-thought prompting, suggesting that GPT-4o is best suited for users who can guide it with more context-aware instructions.

Across all evaluated workflows, the three qualitative dimensions, completeness, correctness, and usability, reveal complementary strengths and limitations among the LLMs. In terms of completeness, Gemini consistently includes all major processing steps and data transformations, especially in Galaxy workflows, often mirroring the structure of official tutorials. DeepSeek also performs well in completeness, particularly for Nextflow, where its outputs capture not only the high-level steps but also underlying configuration details such as profiles and container usage. Correctness is generally high across all models, though some models (especially GPT-4o) occasionally select reasonable but suboptimal tools, or miss platform-specific conventions. Gemini demonstrated strong correctness by adhering to community standards (e.g., ToolShed in Galaxy, nf-core in Nextflow), while DeepSeek shows tool-level precision, particularly in script-based workflows. The greatest variance is observed in usability. Gemini’s outputs are highly readable, modular, and well-suited for instructional or real-world use, especially by users unfamiliar with the workflow system. GPT-4o excels in usability when prompts are structured, though it sometimes lacks operational context. DeepSeek, despite its technical accuracy, often produces verbose and densely structured workflows that could overwhelm novice users, requiring post-processing or simplification for practical adoption. These dimension-wise differences underline the importance of aligning LLM choice with user expertise and platform characteristics.

These results highlight that the effectiveness of an LLM for scientific workflow generation is strongly dependent on the characteristics of the workflow system in use. For graphical environments like Galaxy, where clarity and usability are paramount, Gemini is most effective. For script-based environments like Nextflow, which demand technical completeness and execution logic, DeepSeek performs best. Prompting strategy and model-specific tuning play a critical role in maximizing output quality.
\subsubsection{RQ3: What prompting strategies should a workflow developer follow?}

\textbf{Motivation: }LLMs are sensitive to the way users phrase their requests, and the same task may yield vastly different outcomes depending on the prompting strategy. While LLMs have shown promise in generating scientific workflows, the quality, structure, and reliability of the outputs are not solely functions of the model’s architecture, they are also critically shaped by how the task is presented. For workflow developers, especially those unfamiliar with prompt engineering, this variability introduces uncertainty and inefficiency in practice. Therefore, understanding which prompting strategies, such as instruction-only, role-based, or chain-of-thought, produce more accurate and usable outputs is essential. This RQ investigates how different prompt formulations impact the generation of workflows in terms of correctness, completeness, and usability across models. By identifying effective strategies, we aim to empower developers with practical guidance for interacting with LLMs to support reproducible and efficient scientific workflow development.

\textbf{Approach: }Professionals can achieve various objectives through effective prompts, such as question answering, complex or sequential analysis, and code generation. However, the potential of effective prompting has yet to be fully realized within the workflow community. We recognize the need to demonstrate how effective prompting strategies can empower workflow developers by enhancing automation, improving data handling, and streamlining complex workflows. To address this, we aim to explore existing prompting strategies and identify the most effective approaches for our analysis by exploring the prompting strategies \cite{lin2024write, liu2023jailbreaking, marvin2023prompt, arora2022ask, ekin2023prompt, reynolds2021prompt, wei2022chain, zhou2022large, wang2024prompt, sahoo2024systematic, chen2023unleashing, parameswaran2023revisiting, vatsal2024survey, shah2025prompt, smith2024language, cohn2023towards, razdaibiedina2023progressive,knoth2024ai, lo2023art}. Effective prompting indicates a strong likelihood of receiving the desired recommendations \cite{ekin2023prompt, arora2022ask, wei2022chain, zhou2022large, marvin2023prompt, lin2024write}. In the following, we summarize our strategies to create effective prompting for our work.

\textbf{Results:}
\newline
\textbf{Clear and Specific Instructions: }Providing clear and specific instructions in LLMs is essential for generating the desired output. Ambiguity in the prompt can lead to unexpected responses. For each workflow, we provide precise instructions to get the responses, and most of the time, we receive the desired responses. For example, we ask \emph{What is the purpose of the Galaxy ToolShed?}. This background question is direct and scoped to a specific component of the Galaxy ecosystem. Models like Gemini 2.5 Flash respond with clear definitions and references to community-driven tool sharing, which makes comparison across LLMs reliable.

\textbf{Experimenting with context and examples: }Incorporating well-defined context and relevant examples in prompts can significantly improve the ability of LLMs to generate accurate and domain-specific workflows. Contextualizing prompts ensures that LLMs understand the nuances of the task, while examples serve as guides to steer the model toward producing precise and actionable outputs. This approach is essential when working with specialized domains where generic responses may lack relevance. By providing clear context and concrete examples, we are able to design complex workflows tailored to specific scientific applications. For instance, our prompt, \emph{Create a Galaxy workflow that identifies the exon with the highest number of SNPs on human chromosome 22 using VCF and GTF files} is clear about the biological target (exon with most SNPs), the input file types (VCF, GTF), and the chromosome of interest, helping LLMs produce a complete and executable workflow.

\textbf{Leveraging System 1 and System 2 questions: }To improve the LLMs' responses, understanding the concept of System 1 and System 2 questions is essential. System 1 questions typically involve quick, intuitive, or pattern recognition-based answers. For example, \emph{What does a DAG (Directed Acyclic Graph) represent in scientific workflows?}. On the contrary, System 2 questions involve more deliberate, analytical, complex problem-solving and effortful thinking. For example, \emph{How does the design of a scientific workflow system affect its scalability and performance in large-scale data processing?}

\textbf{Contextual Role Assignment: }Embedding the model in a specific expert role improves technical relevance and response structure. For instance, the prompt \emph{You are a bioinformatics workflow developer with expertise in building reproducible pipelines using Nextflow and nf-core modules. Your task is to design a Nextflow workflow for DNA methylation analysis using bisulfite-converted sequencing (BS-seq) data}, assigns the role so that LLMs can respond accordingly.

\textbf{Controlling Output Verbosity: }Adjusting LLM verbosity tailors responses to the desired detail level. Users can specify response length for concise summaries or detailed explanations, aligning outputs with their needs. Controlling verbosity enhances response relevance and improves workflow design efficiency by matching outputs to the task's context and purpose.

\textbf{User Intent: }Understanding the user's goal and desired output is fundamental to formulating effective prompts. A clear awareness of the intent behind the interaction, whether it involves information retrieval, content generation, or problem-solving, ensures that the prompt aligns with the user's expectations. Users can guide LLMs to produce accurate, relevant, and actionable responses by tailoring the prompt to address the specific objective. This focus on intent-driven prompt design is essential in workflow development, where precision and relevance directly impact workflows quality.

\textbf{Model Understanding: }Understanding the model's knowledge and limitations is essential for designing prompts that maximize its strengths and mitigate weaknesses. Even advanced models like ChatGPT may struggle with certain tasks or produce errors. Awareness of these constraints allows users to compose prompts that align with the model's capabilities. Additionally, incorporating strategies such as providing explicit instructions, clarifying context, or breaking complex tasks into smaller components can help address potential shortcomings. This awareness is vital in workflow development, where precision and domain-specific accuracy are crucial.

\textbf{Domain Specificity: }When working within specialized domains like scientific workflows, providing additional context and carefully chosen examples can significantly improve the robustness and relevance of model-generated responses. Context helps the model understand the specific requirements and constraints of the domain, while examples act as guides to shape the response. This approach ensures that the generated outputs align more closely with domain-specific expectations, enabling the development of precise and actionable scientific workflows.

\textbf{System-Aware Prompting: }Including system-specific terms (e.g., Galaxy or Nextflow) helps models tailor responses to the correct platform. For example, we ask \emph{In the Galaxy platform, how are histories used to manage data?}. The inclusion of \emph{Galaxy} helped models focus their responses specifically on GUI-based data lineage, rather than confusing it with script-based logging mechanisms in other SWSs.

\textbf{Iterative Testing and Refining: }Iterative testing and refinement are key to improving prompt responses. By analyzing outputs and adjusting prompts, users can fine-tune the model for greater accuracy and relevance. This process is especially valuable in workflow development as it enables the gradual optimization of prompts to achieve precise and actionable workflows.

\textbf{Stepwise Reasoning: }Asking models to break down the process logically before generation improves completeness and correctness. Example from Workflow Prompt (Galaxy W4): \emph{First, list the required input files and tools. Then describe each step of the workflow from quality control to mapping. Finally, convert this plan into a Galaxy workflow}. This prompt led  GPT-4o to reason about tool chaining (\emph{assess $\rightarrow$ clean $\rightarrow$ re-assess}) before emitting the final workflow, resulting in better modularity and explanation.

\textbf{Balancing User Intent and LLMs Creativity: }Balancing user intent with LLM creativity is crucial for meaningful responses. While LLMs excel at innovation, prompts must guide outputs to align with user goals. This balance is vital in workflow development, where precision and problem-solving are essential for accurate results.

\textbf{Ensuring ethical usage and avoiding biases: }To ensure ethical use, addressing biases in LLM responses is vital. Clear guidelines, inclusive language, critical evaluation, and content filtering help mitigate issues and prevent harmful stereotypes. Maintaining ethical standards in workflow development ensures that outputs remain unbiased, inclusive, and suitable for diverse audiences, fostering trust and reliability in the model's applications.

\textbf{Prompt Chaining and Multi-turn Conversations: }Prompt chaining connects multiple prompts sequentially, enabling dynamic, context-aware interactions with LLMs. Breaking complex queries into steps facilitates deeper exploration and refined discussions. In workflow development, this approach ensures step-by-step accuracy and cohesive integration of processes.

\textbf{Handling ambiguous or contradictory inputs: }LLMs may occasionally receive ambiguous or contradictory inputs. In such cases, prompt design with clarification is essential. Providing additional context or explicitly stating assumptions can further mitigate confusion. In workflow development, where precision is paramount, designing prompts to handle ambiguity ensures consistent and reliable outputs, even in complex or unclear scenarios.

We also employ several additional prompting techniques to enhance workflow accuracy and completeness. These include feedback loops for error evaluation, progressive prompting to break tasks into steps, backward prompting to verify prior outputs and multi-modal prompts. Together, these strategies refine workflow design and ensure robust results.

\begin{boxK}
Following these strategies, we finalize the prompts for our study, as shown in Table \ref{tab:fundamental_questions_of_SWS_and_SW}, \ref{tab:sws_background_questions}, \ref{tab:galaxy_background_questions}, \ref{tab:nfcore_background_questions}. This table includes prompts designed to gather background information and prior knowledge relevant to our selected workflows. The final workflow-specific prompts are presented in Tables \ref{tab:galaxy_prompts} and \ref{tab:nextflow_prompts}, ensuring a clearer objective of each workflow while minimizing redundancy.
\end{boxK}

To standardize the prompt design process and facilitate reproducible evaluation, we identify generalizable prompt patterns for constructing effective workflow-generation queries in both Galaxy and Nextflow environments. These patterns emerge through a close examination of prompt structures used in our experiments, particularly those presented in Tables \ref{tab:galaxy_prompts} and \ref{tab:nextflow_prompts}.

\begin{boxK}
For Galaxy, which features a graphical user interface and emphasizes accessibility for non-programmers, prompts benefit from highlighting the biological objective, the input file types, and the desired output. A typical prompt for Galaxy follows this pattern: \emph{Create a Galaxy workflow that performs [biological goal or analysis task] using [input file formats, such as VCF, GTF, or FASTQ]. The workflow should include steps for [intermediate processing, such as quality control, filtering, or mapping] and produce [expected output, such as variant calls or expression matrices]}. Prompts constructed using this pattern ensure that the model captures both the domain intent and Galaxy-specific processing logic. For instance, a clear example from our study is: 

\emph{Create a Galaxy workflow that identifies the exon with the highest number of SNPs on human chromosome 22 using VCF and GTF files}, which allowed LLMs to generate workflows with tool-specific operations aligned to GTN tutorials.
    
\end{boxK}

\begin{boxK}
    In contrast, Nextflow is a script-based system designed for high reproducibility and modular pipeline development. Prompts targeting Nextflow must emphasize the computational modules, input-output relationships, and optional configuration logic. An effective prompt pattern in this case is: \emph{Create a Nextflow workflow that performs [bioinformatics task] using [tools, such as fastp, STAR, featureCounts]. The workflow should take [data types] as input and output [result types], including steps for [processes like alignment, quantification, or filtering]}. An example that reflects this pattern is: \emph{Create a Nextflow workflow to process raw FASTQ files using fastp, STAR, and featureCounts, and perform differential expression analysis}, which guides the LLM to structure the script according to Nextflow syntax and pipeline conventions.

These base patterns can be enhanced through prompting strategies such as role-based framing (e.g., \emph{You are a bioinformatics workflow developer using Nextflow}), stepwise reasoning (\emph{First list the inputs and tools, then describe each step before generating the workflow}), and ambiguity handling (\emph{Assume the input data is raw and unfiltered}). Such augmentations help LLMs disambiguate task requirements, infer missing steps, and align their responses more closely with real-world workflows.
\end{boxK}

By articulating these prompt patterns, we provide a foundation for both practitioners and researchers to systematically guide LLMs in generating domain-relevant workflows across different scientific workflow systems.

\section{Discussion} \label{prompt-to-pipeline-discussion}
This study systematically evaluates how state-of-the-art LLMs perform in the development of scientific workflows, with a particular focus on bioinformatics scenarios using Galaxy and Nextflow. Through a task-oriented analysis across three research questions, we examine the capabilities and limitations of LLMs in (i) addressing fundamental and background knowledge of scientific workflow systems, (ii) generating complete and executable workflows, and (iii) responding to different prompting strategies. Our results collectively offer a grounded understanding of the practical value and current constraints of LLMs in computational research workflows.

One of the most compelling findings is that LLMs, particularly Gemini 2.5 Flash and GPT-4o, demonstrate promising performance in addressing fundamental questions about workflow systems. These models can provide syntactically sound and semantically relevant responses to common background queries on workflow components and domain-specific tasks. However, performance variability exists across tools, with some models omitting contextual steps or offering generalized responses, especially when minimal prompting is used. This observation underscores the need for careful prompt design even for tasks that may appear routine to human experts.

When evaluating LLM-generated workflows (RQ2), we uncover significant differences in quality, particularly in the dimensions of completeness, correctness, and usability. Gemini 2.5 Flash consistently produces high-quality workflows that are logically structured, well-annotated, and accessible for novice users. DeepSeek-V3 also performs reasonably well, especially in generating technically detailed responses including command-line and directory structures. However, GPT-4o exhibits inconsistencies when tasked with minimal instruction, often omitting critical setup steps or default configurations. These results suggest that while LLMs can be helpful in automating parts of the workflow development process, their outputs must be critically evaluated, especially in high-stakes scientific analyses.

Our investigation into prompting strategies (RQ3) further illuminates the dependencies between user input structure and model performance. The tiered evaluation, using instruction-only, role-based, and chain-of-thought prompts, revealed that prompt formulation plays a pivotal role in shaping output quality. For simpler tasks, most models respond adequately to concise instructions. However, as the complexity of the task increased, instruction-only prompts are often insufficient. GPT-4o, in particular, demonstrates the most improvement when guided by role-based or step-wise chain-of-thought prompts. In contrast, Gemini generally maintains robust performance even with minimal instruction, suggesting that some models are better tuned for general usability, while others require more deliberate contextual scaffolding to achieve optimal outcomes.

Importantly, the study highlights a practical implication: prompt engineering is not merely an academic curiosity but a crucial skill for practitioners leveraging LLMs in scientific workflow development. While the models exhibit impressive generative capacity, their effectiveness is shaped by the user's ability to structure requests, assess outputs, and iteratively refine instructions. For research groups aiming to integrate LLMs into their scientific pipeline design process, this emphasizes the need to develop domain-specific prompting strategies and validation routines to ensure robustness and reproducibility.

Nevertheless, the study also reveals limitations. LLMs still occasionally produce hallucinated tools or incompatible parameter settings, and they rarely reflect the full diversity of domain-specific conventions. Furthermore, workflow narratives are often limited in their ability to explain the rationale behind tool selection or parameterization, an essential aspect of reproducibility and scientific communication. These issues point to an important area for future work: the integration of LLMs with domain-aware validation engines and curated knowledge bases to improve fidelity and interpretability.

In summary, while LLMs exhibit growing competence in the generation and explanation of scientific workflows, they are not yet fully autonomous solutions. Their utility depends critically on model choice, prompt design, and human oversight. By understanding and addressing these dependencies, workflow developers can better harness the capabilities of LLMs for accelerating computational research while maintaining scientific rigor and transparency.

\section{Threats to Validity} \label{threats-of-prompts-to-pipeline}
Despite the rigorous evaluation conducted in this study, we acknowledge several potential threats to validity to contextualize the scope and applicability of our findings.

\textbf{Internal Validity: }refers to the extent to which a causal relationship can be established between the applied treatment and the observed outcomes \cite{wohlin2012experimentation}. A central threat to the internal validity of our investigation lies in the subjectivity of manual workflow evaluation. While the assessment of correctness, completeness, and usability was conducted by two domain experts with significant experience in scientific workflows, human interpretation is inherently susceptible to bias. To mitigate this, we relied on well-defined rubrics and cross-validation against established baselines from the Galaxy Training Network (GTN) and nf-core repositories. Nevertheless, the absence of quantitative metrics for some usability aspects introduces potential inconsistencies in evaluation outcomes. Additionally, while we adopted a stepwise prompting approach, progressing from instruction-only to role-based and chain-of-thought prompts, our criteria for escalation may still carry implicit judgment calls. In rare cases, the decision not to escalate a prompt may overlook latent model capabilities that might emerge under more nuanced scenarios.

\textbf{ External Validity: }refers to the extent to which the research findings can be generalized beyond the specific context of the study to broader populations, settings, or situations \cite{wohlin2012experimentation}. The generalizability of our findings may be limited by the specific selection of LLMs and workflows. Our study evaluated three LLMs \emph{GPT-4o, Gemini 2.5 Flash, and DeepSeek-V3}, all chosen based on their availability and relevance as of early 2025. As the LLM landscape evolves rapidly, newer models may outperform or behave differently under the same experimental settings. Furthermore, although the selected workflows span a wide range of bioinformatics tasks and platforms (Galaxy and Nextflow), they may not capture the full diversity of scientific workflows used in other domains or emerging subfields within bioinformatics. Our workflow tasks also emphasized well-documented use cases from GTN and nf-core. While this ensures reproducibility and standardization, it may not reflect more exploratory or unconventional workflows encountered in cutting-edge research, where documentation is sparse or evolving.

\textbf{Construct Validity: }concerns the extent to which the measurements and observations accurately reflect the theoretical concepts they are intended to represent \cite{wohlin2012experimentation}. While \emph{completeness, correctness, and usability} are standard in workflow assessment, their practical interpretation may vary across evaluators and contexts. For example, a workflow deemed usable for an expert might not be equally accessible to a novice. Although we attempted to account for varying user expertise levels by noting model behavior across different prompting strategies, our construct definitions could still be refined further in future studies using more granular user studies or empirical execution logs. Furthermore, the assumption that community-curated workflows represent ground truth may overlook acceptable alternative implementations. Some LLM-generated workflows deviated from reference workflows in structure or tool usage but remained valid and executable. While we considered such variation acceptable when logically justified, this introduces a gray area in evaluation that may affect reproducibility across studies.

\textbf{ Conclusion Validity: } refers to the degree to which the conclusions drawn from a study are based on appropriate and reliable analysis of the data. Our conclusions regarding model effectiveness and prompting strategy are based on a relatively small sample size of workflows and limited prompting techniques. While this setup was chosen to balance depth and manageability, the limited scope may affect the statistical power of our inferences. Additionally, since all models were evaluated using their default configurations, we did not explore the impact of parameter tuning or prompt fine-tuning, which might significantly influence output quality.
By recognizing these limitations, we aim to foster transparency and encourage further research to replicate, validate, and extend our findings across broader contexts, user bases, and scientific domains.

By recognizing these limitations, we aim to foster transparency and encourage further research to replicate, validate, and extend our findings across broader contexts, user bases, and scientific domains.

\section{Related Work} \label{related-work-of-prompts-to-pipeline}
Scientific workflow development plays a central role in modern bioinformatics, supporting complex analyses that involve multiple tools/modules, datasets, and computing environments. However, constructing these workflows remains a significant challenge, particularly for researchers without extensive programming expertise. To simplify workflow construction and scientific analysis, researchers have developed several approaches. For instance, Palmblad et al. \cite{palmblad2019automated} used EDAM and semantic tool annotations to enable automated workflow generation in mass spectrometry-based proteomics. Similarly, Di Bernardo et al. \cite{dibernardo2008semi} utilized data types to generate workflows automatically. Koop et al. introduced \emph{VisComplete}, a system designed to help users build visualization workflows based on prior workflows in the VisTrails framework \cite{koop2008viscomplete}. However, this approach did not verify the correctness of the reused workflows. These early methods, while innovative, were limited in scope, applicable to a narrow range of bioinformatics analyses, and prone to errors as tools evolved. Addressing these challenges, Kumar et al. \cite{kumar2021tool} developed a deep learning-based model for recommending tools during workflow construction by analyzing workflows hosted on the European Galaxy server. However, their approach overlooked tool compatibility, failed to account for outdated tools and was restricted to workflows from the European Galaxy server. Building on this, Kumar et al. \cite{kumar2023transformer} developed a transformer-based tool recommendation system, but it lacked generalizability beyond the European Galaxy platform.  Recent advances in LLMs have opened new opportunities for simplifying workflow development through natural language interaction and automated code generation. Several studies have explored the potential of LLMs to support scientific workflow development, highlighting both their potential and limitations. Sänger et al. \cite{sanger2024qualitative} conducted a qualitative assessment of ChatGPT's capabilities in assisting with workflow design across general scientific domains. While their study demonstrated ChatGPT's ability to structure high-level workflows and reduce development complexity, it was limited to a single model and did not address domain-specific accuracy or tool-level fidelity. In contrast, our work conducts a systematic, domain-specific evaluation of three advanced LLMs, GPT-4o, Gemini 2.5 Flash, and DeepSeek-V3, focusing on bioinformatics workflows across two widely adopted scientific workflow systems, Galaxy and Nextflow. Beyond workflow generation, we explicitly assess the role of prompting strategies (instruction-only, role-based, and chain-of-thought), offering a deeper analysis of LLM behavior in structured scientific contexts. Our evaluation is grounded in community-validated baselines (e.g., GTN and nf-core), enabling reproducible comparisons of workflow quality in terms of completeness, correctness, and usability. Pickard et al. \cite{pickard2024language} proposed \emph{BRAD}, a retrieval-augmented agent that integrates LLMs with external databases to support tasks such as question answering and gene enrichment analysis. While BRAD focuses on multi-modal task execution through an agentic interface, our study centers on the generation of end-to-end executable workflows. Similarly, Riffle et al. \cite{riffle2025olaf} introduced \emph{OLAF}, a conversational platform that enables users to execute single-cell analyses via modular LLM-driven interaction. Unlike these agent-based systems, our approach provides a model-agnostic, benchmark-driven comparison of LLMs, emphasizing prompt effectiveness and workflow synthesis across platforms rather than execution or interaction design.
Tang et al. \cite{tang2024biocoder} presented \emph{BioCoder}, a benchmark suite for evaluating LLM performance in generating bioinformatics code snippets in Python and Java. Their work focuses on isolated function-level code generation and uses fuzz-testing to measure correctness. While informative for programming capabilities, BioCoder does not consider workflow-level composition or usability in scientific contexts. Our study moves beyond code synthesis by generating and validating complete workflows that mirror real-world bioinformatics pipelines, thereby addressing aspects of tool chaining, data flow, and reproducibility.

Other notable efforts include the Playbook Workflow Builder (PWB) \cite{clarke2025playbook}, which offers a GUI-based interface for assembling workflows through semantically annotated building blocks, and PROTEUS \cite{ding2024automating}, which applies LLMs for hierarchical planning and hypothesis generation in proteomics. These systems support accessible workflow construction or autonomous task planning, yet they do not evaluate the quality of LLM-generated workflows against executable community standards, nor do they systematically assess the effect of prompt design. Despite these advancements, the integration of LLMs into practical, domain-specific scientific workflow development remains underexplored. Prior work tends to focus either on high-level planning or isolated code generation, lacking a comprehensive analysis that connects prompt design, model behavior, and platform-specific execution. Our study fills this gap by offering a unified evaluation of LLMs for bioinformatics workflow generation, grounded in real platforms, community standards, and empirical metrics.

\section{Conclusion}\label{conclusion-of-prompts-to-pipeline}
This study presents a systematic evaluation of LLMs for generating scientific workflows in bioinformatics, focusing on two widely adopted workflow systems, Galaxy and Nextflow. By analyzing the performance of GPT-4o, Gemini 2.5 Flash, and DeepSeek-V3 across a curated set of bioinformatics tasks, we examined the models’ capabilities in addressing fundamental workflow concepts, producing end-to-end executable pipelines, and responding to various prompting strategies. Our results highlight that LLMs can serve as effective co-developers for scientific workflows, especially when guided by well-structured prompts. Gemini 2.5 Flash consistently performed well with minimal prompting, offering clear, complete, and intuitive outputs, making it particularly suitable for novice users. GPT-4o, while occasionally omitting details in instruction-only prompts, excelled when provided with role-based context, indicating that prompt engineering plays a critical role in leveraging its full potential. DeepSeek-V3 also demonstrated strong technical depth, especially in producing scripts with directory structures and command-line instructions, although its verbosity and inconsistency may hinder usability for some users. Importantly, our study underscores the need for deliberate prompting strategies. Simple prompts often yield incomplete or ambiguous outputs, whereas thoughtfully layered approaches, incorporating context, roles, and logical decomposition, significantly improve workflow generation. These findings are relevant to developers using LLMs for workflow authoring and designing future LLM-integrated tools for scientific research.

In summary, while LLMs have not yet achieved full autonomy in scientific workflow construction, they already offer valuable support for both novice and expert users. Their integration into scientific toolchains can democratize workflow development, reduce entry barriers, and accelerate research reproducibility, provided their use is guided by informed prompting and critical validation. Future work should explore scalable evaluation frameworks, broader domain coverage, and user-centered studies to further solidify the role of LLMs in scientific workflows.

\section*{Supplementary information}
All supplementary files can be obtained using \cite{alam_2025_16416384}

\section*{Acknowledgment}
This research is supported in part by the Natural Sciences and Engineering Research Council of Canada (NSERC), and by the industry-stream NSERC CREATE in Software Analytics Research (SOAR).

\section*{Data availability statement}
All supplementary files can be obtained using \cite{alam_2025_16416384}
\section*{Declarations}

\begin{itemize}
\item Funding: This research is supported in part by the Natural Sciences and Engineering Research Council of Canada (NSERC), and by the industry-stream NSERC CREATE in Software Analytics Research (SOAR)
\item Conflict of interest/Competing interests: We have no conflict of interest.
\item Ethics approval and consent to participate: Not Applicable
\item Consent for publication: Not Applicable
\item Data availability: Our data can be obtained using \cite{alam_2025_16416384} 
\item Materials availability: Our data can be obtained using \cite{alam_2025_16416384} 
\item Code availability Not Applicable
\item Author contribution: Khairul Alam designed and conducted the study, developed the evaluation framework, constructed prompts, carried out the analysis, and evaluated the LLM-generated workflows. He also led the drafting and writing of the manuscript. Banani Roy contributed to the evaluation of the workflows, provided critical insights on prompt engineering and evaluation criteria, reviewed and edited the manuscript, and offered guidance throughout the study.
\end{itemize}

\appendix

 \bibliographystyle{elsarticle-num} 
 \bibliography{elsarticle-template-num}

\end{document}